%% file: main.tex
\documentclass[a4, preprint, amssymb, amsmath,nofootinbib]{revtex4-2}
\usepackage{amsmath}
\usepackage{amsfonts}
\usepackage{amsthm}
\usepackage{graphics}
\usepackage{graphicx}
\usepackage{subfigure}
\usepackage{amssymb}
\usepackage {mathrsfs}
\usepackage{braket}

\usepackage{color}
\usepackage{url}
\usepackage[abs]{overpic}
\usepackage[usenames,dvipsnames]{xcolor}

\def\v{\textbf{v}}

\newcommand{\moh}[1]{\textcolor{black}{#1}}
\newcommand{\tom}[1]{\textcolor{black}{#1}}
\newcommand{\green}[1]{\textcolor{black}{#1}}
\begin{document}

\title{Revealing self-gravity in a Stern-Gerlach Humpty-Dumpty experiment}

\author{Mohamed Hatifi}
\email{hatifi.mohamed@gmail.com}
\affiliation{Quantum Machine Unit, Okinawa Institute of Science and Technology Graduate University, Onna, Okinawa 904-0412, Japan}

\author{Thomas Durt} 
\email{ thomas.durt@centrale-marseille.fr}
\affiliation{Aix Marseille Universit\'e, CNRS, Centrale Marseille, Institut Fresnel UMR 7249, Marseille, France}
\date{\today}
\begin{abstract}
There is no consensus among today’s physicists about how to describe the gravitational interaction properly in a quantum framework.  We propose in this paper an experimental test aimed at revealing the existence of a non-linear self-interaction \`a la Schr\"odinger-Newton (S-N). In this test, a mesoscopic spin 1/2 microsphere is freely falling in a Humpty-Dumpty Stern-Gerlach interferometer. After clarifying the role of the scaling of the interaction in function of the amplitudes of the up and down spin components of the microsphere, it is shown that self-gravity induces a measurable phase shift between them, which paves the way to experimental tests. It is also shown that if we consider two distinct microspheres falling in parallel, the entangling power of the S-N interaction is exactly equal to zero.
\end{abstract}
\maketitle
\section{Introduction} Properly quantizing gravity remains one of the most challenging problems of today's theoretical physics. Recently several experiments were proposed, aimed at testing manifestations of the gravitational interaction in the mesoscopic regime (e.g., through entanglement \cite{bosesg,vedral}). It is worth mentioning, however, that there exists no unanimous agreement about how to quantize gravity even in the Newtonian limit \cite{hua2014,bosecomment}. Here we focus on a mean-field formulation of Einstein's general relativity originally proposed by M{\o}ller \cite{Moeller} and Rosenfeld \cite{Rosenfeld} in which space-time remains classical
 while the material source term in Einstein's equation is the average stress-energy tensor, averaged over quantum degrees of freedom:  \begin{equation} R_{\mu\nu}-{1\over 2} g_{\mu\nu}R={8\pi G\over c^4} \bra{\Psi}\hat T_{\mu\nu}\ket{\Psi},\end{equation}
with $R_{\mu\nu}$ the Ricci tensor, $g_{\mu\nu}$ the space-time metric, $G$ Newton's constant, $c$, the velocity of light and $\hat T$ the stress-energy tensor.  \tom{In other words, matter is treated in the mean-field regime.}
In the Newtonian limit, a self-gravitational interaction is thus likely to be present, which can be expressed, for instance, in the single-particle case, through the (non-linear) S-N equation  \cite{diosi84,penrose96,Jones}
 \\
\begin{align}
{i}\hbar\frac{\partial\Psi(t,{\bf x})}{\partial t}=-\frac{\hbar^2}{2m}\,\Delta\Psi(t,{\bf x})+V_{ext}({\bf x},t)\Psi({\bf x},t)+\int {d}^3 x' |\Psi(t,{\bf x'})|^2 V(|{\bf x -x'}|) \Psi(t,{\bf x}),\label{NS}
\end{align}
\\
where the Newton potential $V(d)=-Gm^2/d$ contributes to the non-linear  S-N potential while $V_{ext}({\bf x},t)$ represents the external, linear, potential. 
 \tom{Equation (\ref{NS}) can be derived from the self-gravitational contribution to the total energy, which is equal to}
\begin{equation}
\int~d^3x\,\int~d^3x'~\vert\Psi(t,{\bf x})\vert^2\,\vert\Psi(t,{\bf x'})\vert^2\,\frac{(-)G\,m^2}{\vert{\bf x}-{\bf x'}\vert}\label{classgrav}
\end{equation}

\tom{In this semi-classical approach, the gravitational field is considered to be classical and, contrary to quantum gravity approaches, when the source of the gravitational field is prepared in a coherent superposition of non-overlapping localized states, the gravitational potential/metrics is not a superposition state of the potentials/metrics associated to these localized states. As we explain in section \ref{analogy}, this also contrasts with QED, which is a major source of inspiration for quantum gravity \cite{hua2014}.}

Now, gravity is known to differ from other fundamental interactions\footnote{In particular, it is not renormalizable, contrary to gauge theories associated to electromagnetic, weak and strong interactions.} and it could be after all that the S-N equation (and its many-particles generalization) is relevant for modeling gravity in the quantum and mesoscopic regimes \cite{Carlip2008,penrose2}. This approach would also contribute to solve the measurement problem and to elucidate the quantum-classical transition \cite{Colin2017,diosi84,penrose96}. Here we propose an experimental test to reveal the existence of a gravitational self-interaction {\it \`a la} S-N. It is directly inspired by two recent proposals \cite{bosesg,vedral} in which two massive spin 1/2 objects, initially prepared in a factorizable spin state, simultaneously move in parallel Humpty-Dumpty Stern-Gerlach interferometers (as in figure \ref{fig1}). 

\begin{figure}
\begin{center}
	\includegraphics[scale=0.05]{sternDouble.pdf}
	\caption{Illustration of the double Humpty-Dumpty Stern-Gerlach experiment of two massive spin $1/2$ objects.}\label{fig1}
\end{center}
\end{figure}

These proposals were conceived in such a way that the gravitational interaction ultimately induces some (in principle measurable) spin entanglement between the two objects. Their goal \tom{was to reveal the quantum nature of gravitation. Note that in references \cite{bosesg,vedral} it was assumed from the beginning that no gravitational self-interaction was present.} In the present paper, we propose on the contrary to test the existence of \tom{a semi-classical interaction {\it \`a la S-N} which, among others, is characterized by the existence of a self-interaction at the single object level.}

\tom{ In this case (unique object, as in figure \ref{stern1nano}), we show that when a unique spin 1/2 mesoscopic particle interacts with itself due to self-gravity this interaction leads to a dephasing between the spin up and spin down wave packets inside the Stern-Gerlach interferometer. This will lead after recombination to a rotation of the spin, which can, in principle, get revealed by spin tomography after the completion of the Humpty-Dumpty experiment. No self-interaction is assumed to be present in quantum gravity approaches, which explains why in our case, one Humpty-Dumpty interferometer only is sufficient, and not two as in the two aforementioned proposals \cite{bosesg,vedral}.}
 
 \tom{In the case of two objects, the S-N model implies the coexistence of self-interaction and also of two-body interaction. The entangling power predicted in the semi-classical approach when two objects fall in parallel along two Humpty-Dumpty interferometers  as plotted in figure \ref{fig1} is also shown in the present paper to drastically differ from the one predicted in the standard approach \cite{bosesg,vedral} (actually it is shown to be equal to zero).}

\tom{The paper is structured as follows. In section \ref{self}, we recall the main features of gravitational self-interaction experienced by a rigid homogeneous sphere. In section \ref{hdnous}, after clarifying the controversial \cite{inhib} role of the scaling of the interaction in function of the amplitudes of the up and down spin components of the microsphere, we apply these properties in order to estimate the phase shift at the output of the Humpty-Dumpty interferometer, after recombination of the up and down spin components. In section \ref{numer}, we present the results of our numerical estimates. The section \ref{disc} is devoted to various discussions and the section \ref{zero} is devoted to studying the entangling power of the semi-classical gravitational interaction. The last section is devoted to conclusions. Extra computations and supplementary material can be found in the appendix.}

\section{Self-interaction of a homogeneous sphere.\label{self}}
\tom{Let us consider a solid, rigid object, of mass $m$, and let us denote $\Psi(t,{\bf x}_{\text{CM}})$ the wave function of its center of mass. The average mass-density at a location ${\bf x'}$ is then equal to the convolution}

\tom{$ \int d^3{\bf x}_{\text{CM}}  |\Psi(t,{\bf x}_{\text{CM}})|^2F({\bf x'}-{\bf x}_{\text{CM}} )$ where $F({\bf x'})$ represents the (positive) mass-density  of the object at a position ${\bf x'}$ when its center of mass is located at ${\bf x}=0$. It is normalized to $m$: $ \int d^3{\bf x'}F({\bf x'})=m$. The gravitational potential created by such a mass-distribution at a location ${\bf x}$ is nothing else than  $\int d^3{\bf x'}\int d^3{\bf x}_{\text{CM}}  |\Psi(t,{\bf x}_{\text{CM}})|^2F({\bf x'}-{\bf x}_{\text{CM}}) (-G/|{\bf x -x'}|))$. Accordingly, the full self-interaction energy is equal  to
 \begin{align}\int d^3{\bf x'}_{\text{CM}}\int d^3{\bf x}_{\text{CM}}|\Psi(t,{\bf x'}_{\text{CM}})|^2 V^\mathrm{eff}({\bf x}_{\text{CM}}-{\bf x'}_{\text{CM}}) 
 |\Psi(t,{\bf x}_{\text{CM}})|^2\label{selfintener}\end{align} with
\begin{align}\nonumber V^\mathrm{eff}({\bf x}_{\text{CM}}-{\bf x'}_{\text{CM}}) =\int d^3{\bf x} \int d^3{\bf x'}F({\bf x'}-{\bf x}_{\text{CM}}) (-G/|{\bf x -x'}|) F({\bf x}-{\bf x'}_{\text{CM}})\\=\int d^3{\bf x} \int d^3{\bf \tilde x'}F({\bf \tilde x'}) (-G/|{\bf x -\tilde x'-{\bf x}_{\text{CM}}}|) F({\bf x}-{\bf x'}_{\text{CM}})\nonumber \\=\int d^3{\bf \tilde x} \int d^3{\bf \tilde x'}F({\bf \tilde x'}) (-G/|{\bf \tilde x -\tilde x'}|) F({\bf \tilde x}+{\bf x}_{\text{CM}}-{\bf x'}_{\text{CM}}),\label{multibody}\end{align} where we introduced the new variables ${\bf \tilde x'}={\bf x'}-{\bf x}_{\text{CM}}$ and ${\bf \tilde x}={\bf x}-{\bf x}_{\text{CM}}$.
For a rigid sphere of radius $R$, we shall consider that the density inside the sphere is homogeneous, in first approximation. The consistency and validity of this approximation are discussed further in the paper (sections \ref{bla} and \ref{nuclear}). Then the effective potential $V^\mathrm{eff}({\bf x}_{\text{CM}}-{\bf x'}_{\text{CM}})$ does depend on the norm of  ${\bf x}_{\text{CM}}-{\bf x'}_{\text{CM}}$ only. Its exact expression reads \cite{Iwe1982,CDW}}\begin{equation}
V^\mathrm{eff}(d) = \frac{Gm^2}{R}~\!\left(-\frac{6}{5}+\frac{1}{2}\left(\frac{d}{R}\right)^2-\frac{3}{16}  \left(\frac{d}{R}\right)^3+\frac{1}{160}\left(\frac{d}{R}\right)^5\right)\label{fullpot}
\end{equation}
if $d\leq 2R$ with $d=|{\bf x}_{\text{CM}}-{\bf x}'_{\text{CM}}|$; otherwise for larger distances ($d$ larger than twice the radius $R$), one can integrate the internal contributions using Gauss's theorem:
\begin{equation}
V^\mathrm{eff}(d) =- \frac{Gm^2}{d}\quad (d\geq 2R).\label{fullpot2}
\end{equation}
It follows that the center of mass wavefunction (CMWF) will be solution of: 
\\
\begin{eqnarray}\small
\begin{aligned}
{i}\hbar\frac{\partial\Psi(t,{\bf x}_{\text{CM}})}{\partial t}&=-\frac{\hbar^2}{2m}\,\Delta\Psi(t,{\bf x}_{\text{CM}})
+V_{ext}({\bf x}_{\text{CM}},t)\Psi({\bf x}_{\text{CM}},t)& \\
&+\int {d}^3{\bf x'}_{\text{CM}}|\Psi(t,{\bf x'}_{\text{CM}})|^2V^\mathrm{eff}(|{\bf x}_{\text{CM}}-{\bf x}'_{\text{CM}}|)\Psi(t,{\bf x}_{\text{CM}})\label{fullpotNS},
\end{aligned}
\end{eqnarray} 
where $V^\mathrm{eff}(|{\bf x}_{\text{CM}}-{\bf x}'_{\text{CM}}|)$ has been defined in equations (\ref{fullpot},\ref{fullpot2}).
In particular, in what follows, we will consider the limit where the wave function of the center of mass $\Psi(t,{\bf x}_{\text{CM}})$ is peaked with a width small compared to the radius $R$. In that case the effective potential defined in \eqref{fullpot} can be considered as quadratic (see also \cite{chen}):
\begin{equation}
V^\mathrm{eff}(d) \sim \frac{G\,m^2}{R}~\!\left(-\frac{6}{5}+\frac{1}{2}\left(\frac{d}{R}\right)^2\right)\label{nearfield}
\end{equation}
hence when $d<< 2R$ equation \eqref{fullpotNS} takes the form 
\begin{eqnarray}\small
\begin{aligned}
{i}\hbar\frac{\partial\Psi(t,{\bf x}_{\text{CM}})}{\partial t}&=-\frac{\hbar^2}{2m}\,\Delta\Psi(t,{\bf x}_{\text{CM}})
+V_{ext}({\bf x}_{\text{CM}},t)\Psi({\bf x}_{\text{CM}},t) \\
&+\left[\frac{m\omega_{s}^2}{2}\,\left( {\bf x}_{\text{CM}}-\langle {\bf x}_{\text{CM}}\rangle\right)^2+\frac{m\omega_{s}^2}{2}\,\mathcal{Q}(t)-\frac{6}{5}\frac{G\,m^2}{R}\right]\,\Psi(t,{\bf x_{\text{CM}}})\label{fullpotNS2},
\end{aligned}
\end{eqnarray} 
 where $\mathcal{Q}=\langle {\bf x}_{\text{CM}}^2\rangle-\langle {\bf x}_{\text{CM}}\rangle^2$ is the quantum spread in position while the pulsation of the (comoving) harmonic potential $\omega_{s}$  is equal to $\sqrt{\frac{G\,m}{R^3}}=\sqrt{G\,\rho_{sphere}}$, assumed to be constant. It is worth noting here that, as is shown in sections \ref{bla} and \ref{nuclear}, corrections due to the presence of the nuclei which render the mass distribution in the sphere inhomogeneous \cite{chen,CDW} are negligible in the present context.

\section{Humpty-Dumpty Stern Gerlach experiment as a test for the Schr\"odinger-Newton equation}\label{hdnous}
\subsection{Self-interaction of a spin 1/2 rigid sphere.\label{previous}}In a Stern-Gerlach Humpty-Dumpty experiment, a neutral spin 1/2 object passes through two successive Stern-Gerlach devices, as plotted in figure \ref{stern1nano}. In the first magnetic region, the wave packet gets split into its up and down spin components, and it recombines in the second magnetic region. In between those regions, a phase shift accumulates which can be in principle measured after recombination, for instance, by Stern-Gerlach spin tomography.

\tom{The wave function is now a spinorial wave function: }

$\Psi(t,{\bf x}_{\text{CM}})=\beta_+\Psi_+(t,{\bf x'}_{\text{CM}})\ket{+}+\beta_-\Psi_-(t,{\bf x'}_{\text{CM}})\ket{-}$ \tom{where $\Psi_\pm(t,{\bf x'}_{\text{CM}})$ is the (properly normalized to unity) spatial wave function associated to the spin $\pm$ component and $\vert \beta_+\vert^2+\vert \beta_-\vert^2=1$.}

When there is no overlap between the spin components, 

$|\Psi(t,{\bf x}_{\text{CM}})|^2=|\beta_+\Psi_+(t,{\bf x}_{\text{CM}})|^2+|\beta_-\Psi_-(t,{\bf x}_{\text{CM}}|^2$ so that the full energy of self-interaction (\ref{selfintener}) now reads
\begin{align}\int d^3{\bf x'}_{\text{CM}}\int d^3{\bf x}_{\text{CM}}&|\beta_+\Psi_+(t,{\bf x'}_{\text{CM}})|^2 V^\mathrm{eff}({\bf x}_{\text{CM}}-{\bf x'}_{\text{CM}}) 
 |\beta_+\Psi_+(t,{\bf x}_{\text{CM}})|^2\nonumber\\+&|\beta_+\Psi_+(t,{\bf x'}_{\text{CM}})|^2 V^\mathrm{eff}({\bf x}_{\text{CM}}-{\bf x'}_{\text{CM}}) 
 |\beta_-\Psi_-(t,{\bf x}_{\text{CM}})|^2\nonumber\\+&|\beta_-\Psi_-(t,{\bf x'}_{\text{CM}})|^2 V^\mathrm{eff}({\bf x}_{\text{CM}}-{\bf x'}_{\text{CM}}) 
 |\beta_+\Psi_+(t,{\bf x}_{\text{CM}})|^2\nonumber\\+&|\beta_-\Psi_-(t,{\bf x'}_{\text{CM}})|^2 V^\mathrm{eff}({\bf x}_{\text{CM}}-{\bf x'}_{\text{CM}}) 
 |\beta_-\Psi_-(t,{\bf x}_{\text{CM}})|^2\end{align} 
The (effective) potential generated by such a superposition state  is the sum of 2 potentials:
 \begin{align}\int d^3{\bf x'}_{\text{CM}}|\beta_+\Psi_+(t,{\bf x'}_{\text{CM}})|^2 V^\mathrm{eff}({\bf x}_{\text{CM}}-{\bf x'}_{\text{CM}}), \text{and}\label{previouseqn1}\\\int d^3{\bf x'}_{\text{CM}}|\beta_-\Psi_-(t,{\bf x'}_{\text{CM}})|^2 V^\mathrm{eff}({\bf x}_{\text{CM}}-{\bf x'}_{\text{CM}}),\label{previouseqn2}\end{align} 
respectively generated by the masses distributions $ \int d^3{\bf x}_{\text{CM}}  |\beta_+\Psi_+(t,{\bf x}_{\text{CM}})|^2F({\bf x'}-{\bf x}_{\text{CM}} )$ and $ \int d^3{\bf x}_{\text{CM}}  |\beta_-\Psi_-(t,{\bf x}_{\text{CM}})|^2F({\bf x'}-{\bf x}_{\text{CM}} )$, which are normalised to $|\beta_+|^2\cdot m$ and $|\beta_-|^2\cdot m$
 
  \tom{The potential scales like $|\beta_\pm|^2$, in contrast  (as we explain in sections \ref{previous} and \ref{analogy}) with what could be expected from an analogy with QED \cite{hua2014}, and, not amazingly, with the predictions made in references \cite{bosesg,vedral}, where the interactions scale like $m$ (or $m|\beta_\pm|^0$). It is important to distinguish these various scalings in order to avoid deep misunderstandings as in the treatment of the Humpty-Dumpty interferometer developed in Ref.\cite{inhib}. In that paper, one can read the following comment about an earlier presentation \cite{arxiv} of some of the results presented here: {\it...In a recent preprint, Hatifi and Durt pursue the same idea of a test of the SN equation in a Stern-Gerlach interferometer as described by Bose et al.. (...) their derivation seems fundamentally flaw...The intuition between the considerations in reference \cite{arxiv} seems to be a mental split of the wave function into two particles of masses $|\alpha|^2m$ and $|\beta|^2m$, respectively....} \green{Obviously, there is no mental construction in our derivation which is a straightforward consequence of equation (\ref{selfintener}). Moreover, if the analogy with two particles of masses $|\alpha|^2m$ and $|\beta|^2m$ would be exact, the scaling would be quartic in the moduli of $\beta_\pm$, and not quadratic...}}
  \subsection{Single Humpty-Dumpty Stern-Gerlach experiment}
\begin{figure}[h!]
\begin{center}
\includegraphics[scale=0.045]{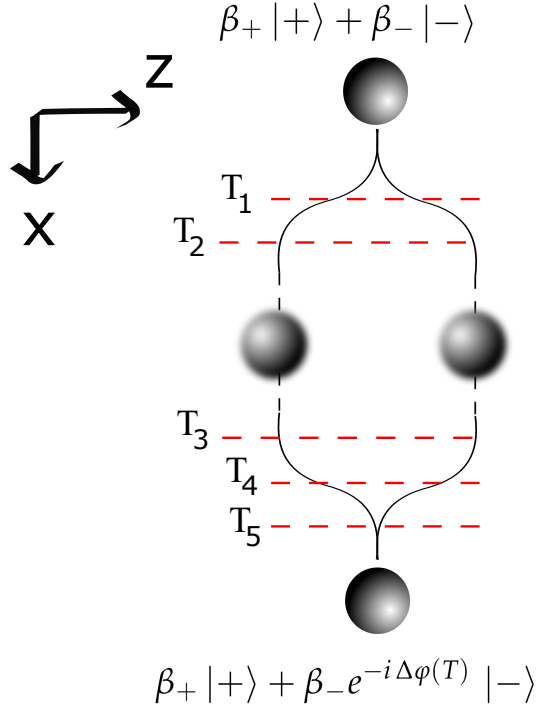}  
\caption{\small Illustration of the (single) Humpty-Dumpty Stern-Gerlach experiment \cite{humpty}. We consider here a freely falling mesoscopic sphere (spin 1/2 NV center in a diamond nanocrystal) of radius $R=1\cdot 10^{-6}$ m and with a mass $m=5.5\cdot 10^{-15}$ Kg. }\label{stern1nano}
\end{center}
\end{figure}
We will consider here a Humpty-Dumpty experiment similar to the one considered in \cite{bosesg}: for instance, a micro-diamond with an embedded NV center spin is released from an optical trap of frequency 1 Mhz, after which it falls freely and directly enters a Stern-Gerlach apparatus where it undergoes, at well-chosen times, a combination of judiciously chosen operations (as e.g. $\pi/2$ spin flips, swaps between electronic states and nuclear spin state) which are described with great detail in \cite{bosesg}. The magnetic field is thus parallel to $Z$, and its amplitude is equal to $B_0-B'_0\,z$, with $B_0$ is the magnetic field and $B'_0$ is its gradient, while the parameter $\lambda$ depends on the branch of the evolution (see figure \ref{stern1nano}) and is defined as follows:
\\
\begin{equation}
\label{lambdat}
\lambda=\left\{ \begin{array}{rcl}
\small &1&\quad\text{if}\quad 0\leq t\leq T_1 \quad \text{or}\quad T_4\leq t\leq T_5  \\
&0&\quad\text{if}\quad T_2\leq t\leq T_3\\
&-1&\quad\text{if}\quad T_1\leq t\leq T_2 \quad\text{or}\quad T_3\leq t\leq T_4
\end{array}\right.
\end{equation}
\\
 A difference with the proposal \cite{bosesg} is that we shall assume that we prepare the center of mass wave function in the ground state of the trap; the reason, therefore, is that all our computations are based on the assumption that the initial state is a gaussian state \cite{CDW}. 

\subsection{Temporal evolution\label{temporal}}

According to the analysis performed in section (\ref{previous}), $V^G$ is the self-gravitational potential defined in \eqref{fullpotNS2} rewritten with convenient weightings for the two spin components (in conformity with equations (\ref{previouseqn1},\ref{previouseqn2})).  If the distance between the centers of the two spin components is larger than twice the radius of the nanosphere ($d_\pm \geq 2R$),  we shall neglect the overlap between the wave packets associated to the centers of mass of the spin components, which is actually fully justified in the narrow wave packet regime where we operate.  Now, with the spin components falling in parallel, the condition $d_\pm=\vert \langle  z \rangle_+ -\langle  z \rangle_-\vert\leq 2R$ is satisfied when $0\leq t \leq T_s$, and $T_5-T_s\leq t \leq T_5$ where the separation (recombination) time $T_s$ can be shown (\ref{sep}) to obey $T_s=\left(\frac{4\,mR}{g\mu_B\,B'_0}\right)^{1/2}$.  Actually, the separation (recombination) time is very short ($T_s\sim$ 0.034 s in the regime considered here, with a free fall of the order of 1 s) so that most of the time the two components do not overlap.
  The potential, at the location of the $\pm$ spin component, is then the sum of the ``near field'' (\ref{nearfield}) generated by the same spin component and the ``far field'' (\ref{fullpot2}) generated by the other spin component. The evolution equation (\ref{fullpotNS2}) thus takes the form
 
 \begin{eqnarray}\small
\begin{aligned}
&{i}\hbar\frac{\partial\Psi_\pm(t,{\bf x}_{\text{CM}})}{\partial t}=-\frac{\hbar^2}{2m}\,\Delta\Psi_\pm(t,{\bf x}_{\text{CM}})
+V_{ext}({\bf x}_{\text{CM}},t)\Psi_\pm({\bf x}_{\text{CM}},t) \\
&+\left[|\beta_\pm|^2(\frac{m\omega_{s}^2}{2}\,\left( {\bf x}_{\text{CM}}-\langle {\bf x}_{\text{CM}}\rangle\right)^2+\frac{m\omega_{s}^2}{2}\,\mathcal{Q}(t)-\frac{6}{5}\frac{G\,m^2}{R})+\vert \beta_\mp \vert^2\,f_\mp (t,{\bf x_{\text{CM}}})\right]\,\Psi_\pm(t,{\bf x_{\text{CM}}})\label{fullpotNS3},
\end{aligned}
\end{eqnarray} 

where $f_\mp (t,{\bf x_{\text{CM}}})$ represents the effective potential between different spin components.

 Actually, in the interval $[T_s,T_5-T_s]$, the Newton force between the up and down wave packets can be shown to be negligibly small. It is easy to check indeed that even if the spin up and down components would move side by side ($d_\pm \approx 2R$) during a time of the order of $T_5$ their Newtonian attraction is so weak that it would reduce the distance between the wave packets by a tiny fraction (10$^{-6}$) of their size $R$.
 Consequently, we shall limit ourselves to the lowest order in the Taylor development of $- \frac{Gm^2}{|z_\pm-<z_\mp>|}$ around $<z_\pm>$ which means that we approximate $f_\mp$ through a classical Newton-like potential: 
  \begin{equation}\label{newpotsg}
f_\mp (t)=-{G\,m^2}\,\frac1{\vert \langle  z \rangle_+ -\langle  z \rangle_-\vert}=-{G\,m^2}\,\frac1{d_\pm}
\end{equation}
Then $f_\mp (x,y,z,t)$ takes the form of a Newton-like potential  ($- \frac{Gm^2}{|z_\pm-<z_\mp>|}$ as in (\ref{fullpot2})); moreover it does no longer depend on $x$ and $y$.

Even inside magnetic regions of the Stern-Gerlach device, equation  (\ref{fullpotNS3}) is thus separable in Cartesian coordinates. This allows us to consider in what follows a description in the  (freely falling) comoving frame in which we limit our study to the component of the CMWF along the quantization axis ($Z$) of the Stern-Gerlach interferometer (see figure \ref{stern1nano}). Nothing remarkable happens along the free fall axis $X$ or along the third axis $Y$: in good approximation, the evolution is the same as for a freely expanding wave packet as shown in appendix (section \ref{sgap1}). After factoring out its $x$ and $y$ components, the CMWF can be expressed as a superposition of spin up and down along $Z$ states
$\large
\Psi(z,t)=\sum_{i=\{+,-\}}\,\beta_i\,\psi_i(z,t)\,\ket{i}\label{supsg}
$. The probability density assigned to the projection of the position of the center of mass along $Z$ reads\begin{equation}
\vert\Psi(z,t)\vert^2=\sum_{i=\{+,-\}}\,\vert \beta_i\vert^2\,\vert\psi_i(z,t)\vert^2,
\end{equation}
with $\vert \beta_+\vert^2+\vert \beta_-\vert^2=\int_{-\infty}^{+\infty}dz |\psi_i(z,t)|^2=1$, and the evolution of the projection along $Z$ of each spin component is described by a Hamiltonian of the form

\begin{equation}\label{hamilsg}
\mathcal{H}_z=-\frac{\hbar^2}{2\,m}\,\frac{\partial^2}{\partial z^2}+\lambda\,\frac{g\mu_B}{2}\left(B_0-B'_0\,z\right)\otimes\sigma_z+V^G(z,t)
\end{equation}
where $V^G$ represents the gravitational self-interaction, $\mu_B$ is the Bohr magneton, $g\sim 2$ is the electronic g-factor and $\lambda$ obeys equation (\ref{lambdat}).

After developing the ``projected''S-N potential around $\langle  z \rangle_+$ (resp. $\langle  z \rangle_-$) we get
 \\
{
\begin{equation}
V_{\pm}^G(z,t)=\underbrace{\vert \beta_\pm \vert^2\,\left[\frac{m}{2}\,\omega_{s}^2\left(z-\langle  z \rangle_\pm\right)^2+\frac{m\omega_{s}^2}{2}\,\mathcal{Q}_\pm(t)-\frac{6}{5}\,\frac{G\,m^2}{R}\right]}_\textrm{\small Self-interaction $\ket{\pm}\rightarrow \ket{\pm}$}\nonumber+\underbrace{\vert \beta_\mp \vert^2\,f_\mp (z,t)}_\textrm{\small Self-interaction $\ket{\mp}\rightarrow \ket{\pm}$}\label{vgpmone}
\end{equation}
 }
 \\
 where $\mathcal{Q}_\pm=\langle z^2\rangle_\pm-\langle z\rangle_\pm^2$, and the function $f_\mp (z,t)=\int {d} z'~ |\psi_\mp(t,{\bf z'})|^2 ~V^{eff}(|{\bf z -z'}|)
$. \tom{We made use of the fact that $V^{eff}(|{\bf z -z'}|)$ interpolates, according to (\ref{fullpot}), between the harmonic potential when $d_\pm=\vert \langle  z \rangle_+ -\langle  z \rangle_-\vert\leq 2R$ and the Newtonian potential if $d_\pm \geq 2R$, as in (\ref{fullpot2}); it also obeys a scaling quadratic in $|\beta_\pm|$.}

 To conclude, each branch of the superposition is solution of the following non-linear Schr\"odinger equation:
\begin{widetext}
\begin{equation}
i\hbar\,\frac{\partial\psi_\pm(z,t)}{\partial t}=\left[-\frac{\hbar^2}{2\,m}\,\frac{\partial^2}{\partial z^2}\pm\lambda\frac{g\mu_B}{2}\left(B_0-B'_0\,z\right)+V_{\pm}^G(z,t)\right]\,\psi_\pm(z,t),\label{supsch}
\end{equation}
\end{widetext}
where in good approximation and most of the time $V_{\pm}^G(z,t)$ is the sum of a quadratic self-interaction of each packet with itself (with weight $|\beta_i|^2 $) with an effective Newtonian interaction towards the other wave packet (with weight $|\beta_{j,j\not= i}|^2 $).  During the (very short) separation and recombination periods, the self-gravitational interaction does not contribute much to the dephasing, because the separation (recombination) time is very short ($T_s\sim$ 0.034 s in the regime considered here, with a free fall of the order of 1 s) and also because during the (short) separation (recombination) process, the Stern-Gerlach potential \cite{hsu2011,platt} dominates the self-interaction. We shall thus simply do as if each spin component was isolated during that period.

In accordance with the previous discussion, we have thus
\begin{align}\label{vgpm}
V_{\pm}^G(z,t)=\nu_\pm^2\,\left[\frac{m}{2}\,\omega_{s}^2\left(z-\langle  z \rangle_\pm\right)^2+\frac{m\omega_{s}^2}{2}\,\mathcal{Q}_\pm(t)-\frac{6}{5}\,\frac{G\,m^2}{R}\right]-(1-\nu_\pm^2)\,\frac{Gm^2}{d_\pm}
\end{align}
where
\begin{align}
 \quad \nu_\pm=\left\{ \begin{array}{rcl}
\small &1&\quad\text{if}\quad d_\pm=\vert \langle  z \rangle_+ -\langle  z \rangle_-\vert \leq 2R \\
&\vert \beta_\pm\vert&\quad\text{otherwise that is to say when $T_s\leq t \leq T_5-T_s$ with $T_s=\left(\frac{4\,mR}{g\mu_B\,B'_0}\right)^{1/2}\sim 0.034 s$}
\end{array}\right.
\end{align}
Such an evolution being gaussian, gaussian wave packets remain so in good approximation during the temporal evolution, which seriously facilitates the numerical treatment.
\subsection{{\bf Ehrenfest's theorem and self-gravity}}\label{selfehr}

Before we compute the phase shift associated to each wave packet, it appears to be useful to develop the potential $V(z,t)$ to the second order in $z$ around the location of the peak of the gaussian wave packet. To see this, let us consider the Schr\"odinger equation

\begin{equation}
i\hbar\frac{\partial\Psi}{\partial t} = -\frac{\hbar^{2}}{2m}\frac{\partial^{2}\Psi}{\partial z^{2}} + V(\braket{z}) + \frac{\partial V}{\partial z}\Bigr|_{\braket{z}}(z-\braket{z})+  \frac{1}{2}\frac{\partial^2 V}{\partial z^2}\Bigr|_{\braket{z}}(z-\braket{z})^2
\end{equation}
Denoting $\frac{\partial^2 V}{\partial z^2}\Bigr|_{\braket{z}}=m\omega^2$, and identifying $V_{0}(t)+V_{1}(t)\,z+V_{2}(t)\,z^2$ with the expression of $V$ above we get
\begin{align}
V_0&=V(\braket{z})-\frac{\partial V}{\partial z}\Bigr|_{\braket{z}}\braket{z}+\frac{m\omega^2}{2}\braket{z}^2\\ \nonumber \\
V_1&=\frac{\partial V}{\partial z}\Bigr|_{\braket{z}}-m\omega^2\braket{z} \\  \nonumber \\
V_2&=\frac{m\omega^2}{2}
\end{align}
Generally, a Schr\"odinger equation of the form 
\begin{equation}
i\hbar\,\frac{\partial\psi_\pm(z,t)}{\partial t}=\left[-\frac{\hbar^2}{2\,m}\,\frac{\partial^2}{\partial z^2}+V_{0,\pm}(t)+V_{1,\pm}(t)\,z+V_{2,\pm}(t)\,z^2\right]\,\psi_\pm(z,t)
\end{equation}
can be solved using a gaussian wave function
\begin{equation}
\psi_\pm(z,t)=\exp\left[-A_\pm(t) \frac{z^2}{2}+B_\pm(t)z+C_\pm(t)\right]\label{gauss}
\end{equation} where $A_\pm(t)$, $B_\pm(t)$ and $C_\pm(t)$ are complex functions of time. We then get, after straightforward computations, the following system of equations:\\ \\
\begin{equation}\large
\left\{ \begin{array}{rcl}
i\,\frac{d A_\pm(t)}{dt}&=&\frac{\hbar}{m}\,A_\pm(t)^2-2\,\frac{V_{2,\pm}(t)}{\hbar}\\ \\
i\,\frac{d B_\pm(t)}{dt}&=&\frac{\hbar}{m}\,A_\pm(t)\,B_\pm(t)+\frac{V_{1,\pm}(t)}{\hbar}\\ \\
i\,\frac{d C_\pm(t)}{dt}&=&\frac{\hbar}{2\,m}\,\left[A_\pm(t)-B_\pm(t)^2\right]+\,\frac{V_{0,\pm}(t)}{\hbar}\,.
\end{array}\right.\label{gausg}
\end{equation} \\ \\
where $V_{k,\pm}$, $k=0,1,2$, is defined through \eqref{hamilsg} and \eqref{vgpm}\\
A gaussian packet of the form (\ref{gauss}) is characterized by the following identities:
\begin{equation}
\langle  z\rangle_\pm=\frac{ \mathcal{R}\text{e}B_\pm}{ \mathcal{R}\text{e}A_\pm}\quad\text{and}\quad \langle  p\rangle_\pm=\hbar\,\left(\mathcal{I}\text{m}B_\pm-\mathcal{I}\text{m}A_\pm\,\frac{\mathcal{R}\text{e}B_\pm}{\mathcal{R}\text{e}A_\pm}\, \right)
\end{equation}
And we also know from (\ref{gausg}) that
\begin{align}\small
\frac{d \mathcal{R}\text{e}A_\pm}{dt}&=2\,\frac{\hbar}{m}\,\mathcal{R}\text{e}A_\pm\,\mathcal{I}\text{m}A_\pm, &\frac{d \mathcal{I}\text{m}A_\pm}{dt}&=-{\hbar\over m}(\mathcal{R}\text{e}A_\pm^2- \mathcal{I}\text{m}A_\pm^2)+2\,{V_{2,\pm}\over \hbar} \\\nonumber 
\frac{d \mathcal{R}\text{e}B_\pm}{dt}&={\hbar\over m}\,(\mathcal{R}\text{e}A_\pm\,\mathcal{I}\text{m}B_\pm+\mathcal{I}\text{m}A_\pm\,\mathcal{R}\text{e}B_\pm),&\frac{d \mathcal{I}\text{m}B_\pm}{dt}&=-{\hbar\over m}(\mathcal{R}\text{e}A_\pm\,\mathcal{R}\text{e}B_\pm- \mathcal{I}\text{m}A_\pm\,\mathcal{I}\text{m}B_\pm)-{V_{1,\pm}\over\hbar}
\label{DNL}\end{align}
hence one can show the following identities
\begin{align}
\frac{d\langle  z\rangle_\pm}{dt}&=\frac{1}{m}\,\langle  p\rangle_\pm&\frac{d\langle  p\rangle_\pm}{dt}&=-(V_{1,\pm}+2\,V_{2,\pm}\,\langle  z\rangle_\pm\,)
\end{align}
In particular, (\ref{vgpm}) imposes that $V^G_{1,\pm}=-m\,\omega_{s}^2\nu_\pm^2\,\langle  z\rangle_\pm$ and $V^G_{2,\pm}=\frac{m\omega_{s}^2}{2}$. It follows that $V^G_{1,\pm}+2\,V^G_{2,\pm}\,\langle  z\rangle_\pm=0$ so that
\begin{align}
\frac{d\langle  z\rangle_\pm}{dt}&=\frac{1}{m}\,\langle  p\rangle_\pm&\frac{d\langle  p\rangle_\pm}{dt}&=\pm\lambda\frac{g\mu_B}{2}\,B'_0
\end{align}
\\
Consequently, the trajectory center of mass is not affected by self-gravity, which is a very general feature of S-N equation\footnote{The absence of self-acceleration can be attributed to Galilei invariance as has been shown in Ref.\cite{hatifi2018}, making use of Ehrenfest's theorem. Galilei invariance itself results, via Noether's theorem, from the fact that S-N interaction is invariant under translations.}; it is only influenced by magnetic forces. This allows us to express analytically the average values of the position and that of the momentum of the center of mass which obey the classical predictions at all times (as we do now).
\subsection{Time evolution of average values of the position and momentum \label{sgap2}}
\begin{figure}
\begin{center}
	\includegraphics[scale=0.535]{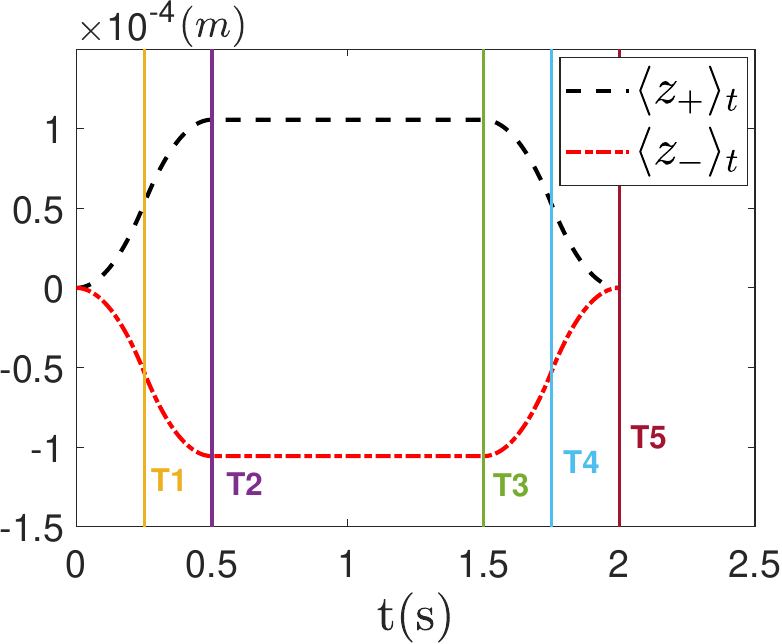}
	\includegraphics[scale=0.55]{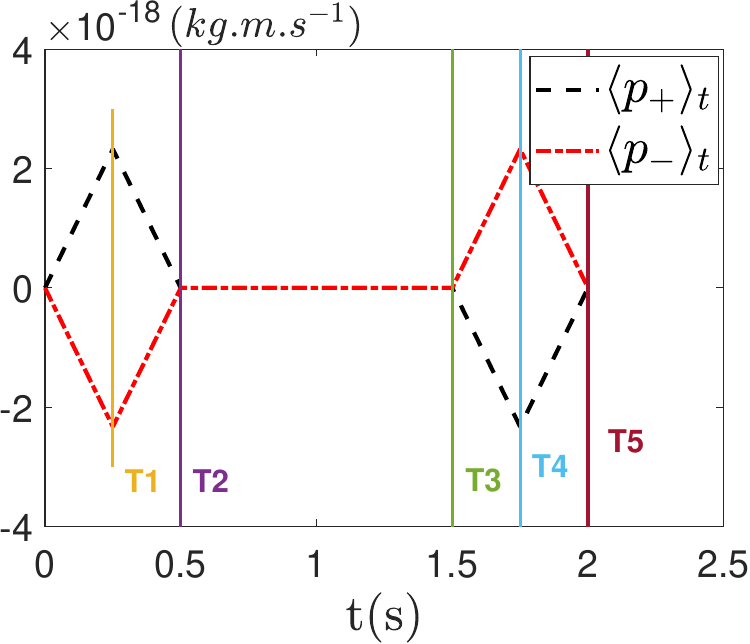}
	\caption{We illustrate here the time evolution of $\langle  z \rangle_\pm$ (left) and $\langle  p \rangle_\pm$ (right), for the same nanosphere as in figure \ref{stern1nano}, with a field gradient $B_0'=10^{6}$ T.m$^{-1}$.}\label{sghp}
\end{center}
\end{figure}

$\bullet\quad T_0 \parbox{1.5cm}{\rightarrowfill} T_1$
\begin{equation}
\langle  p\rangle_\pm =\pm\frac{g\mu_B}{2}\,B'_0\,t\quad\text{and}\quad\langle  z\rangle_\pm =\pm\frac{g\mu_B}{4\,m}\,B'_0\,t^2\label{sep}
\end{equation}
\\
$\bullet\quad T_1 \parbox{1.5cm}{\rightarrowfill} T_2$
\begin{equation}
\langle  p\rangle_\pm =\mp\frac{g\mu_B}{2}\,B'_0\,\left(t-2\,T_1\right)\quad\text{and}\quad\langle  z\rangle_\pm =\mp\frac{g\mu_B}{4\,m}\,B'_0\,\left(t^2-4\,T_1\,t+2\,T_1^2\right)
\end{equation}
\\
$\bullet\quad T_2 \parbox{1.5cm}{\rightarrowfill} T_3$
\begin{equation}
\langle  p\rangle_\pm =0\quad\text{and}\quad\langle  z\rangle_\pm =\pm\frac{g\mu_B}{2\,m}\,B'_0\,T_1^2
\end{equation}
\\
$\bullet\quad T_3 \parbox{1.5cm}{\rightarrowfill} T_4$
\begin{equation}
\langle  p\rangle_\pm =\mp\frac{g\mu_B}{2}\,B'_0\,\left(t-T_3\right)\quad\text{and}\quad\langle  z\rangle_\pm =\mp\frac{g\mu_B}{4\,m}\,B'_0\,\left[\left(t-T_3\right)^2-2\,T_1^2\right]
\end{equation}
\\
$\bullet\quad T_4 \parbox{1.5cm}{\rightarrowfill} T_5$ 
\begin{equation}
\langle  p\rangle_\pm =\pm\frac{g\mu_B}{2}\,B'_0\,\left(t-T_5\right)\quad\text{and}\quad\langle  z\rangle_\pm =\pm\frac{g\mu_B}{4\,m}\,B'_0\,\left(t-T_5\right)^2
\end{equation}
Note that in order to achieve the recombination we must require that
\begin{eqnarray}
 T_2-T_1=T_4-T_3=T_5-T_4=T_1
\label{tint}
\end{eqnarray}
In the numerical simulations, we chose $T_1=0.25\text{ s}$ and $T_3-T_2=1 \text{ s}$ (figure \ref{sghp}).

\subsection{{\bf Classical action and quantum contributions to the phase shifts}}
Let us now compute the phase shift associated with each wave packet. Imposing a gaussian solution (\ref{gauss}) we find, making use of (\ref{gausg}) that the phase $\mathcal{I}\text{m}C$ evolves in time according to
\begin{equation}\label{c2sg}
\frac{d \mathcal{I}\text{m}C}{dt}=\frac{\hbar}{2\,m}\,\left[ (\mathcal{R}\text{e}B)^2-(\mathcal{I}\text{m}B)^2-\mathcal{R}\text{e}A\right]-\,\frac{V_{0}}{\hbar}
\end{equation}
with
\begin{equation}
V_{0}=V(\braket{z}) - \braket{z} \frac{\partial V}{\partial z}\Bigr|_{\braket{z}}+\frac{1}{2}\frac{\partial^2 V}{\partial z^2}\Bigr|_{\braket{z}}\braket{z}^2
\end{equation}
Noting that 
\begin{equation}\label{ehrenf}
\langle z \rangle=\frac{ \mathcal{R}\text{e}B}{ \mathcal{R}\text{e}A}\quad\text{and}\quad \langle p \rangle=\hbar\,\left(\mathcal{I}\text{m}B-\mathcal{I}\text{m}A\,\langle z \rangle \right)
\end{equation}
on can show that\\
\begin{align}
\frac{d \mathcal{I}\text{m}C}{dt}&=\frac{\hbar}{2\,m}\,\left[\langle z \rangle^2\left[(\mathcal{R}\text{e}A)^2-(\mathcal{I}\text{m}A)^2\right]-2\,\frac{\langle p\rangle\,\langle z\rangle}{\hbar}\,\mathcal{I}\text{m}A -\frac{\langle p\rangle^2}{\hbar^2}- \mathcal{R}\text{e}A\right] \nonumber\\ 
&-\,\frac{1}{\hbar}\left[ V(\braket{z}) - \braket{z} \frac{\partial V}{\partial z}\Bigr|_{\braket{z}}+\frac{1}{2}\frac{\partial^2 V}{\partial z^2}\Bigr|_{\braket{z}}\braket{z}^2\right]
\end{align}
Making use of Ehrenfest's theorem, we get, as already shown $ \frac{d \braket{p}}{dt}=-\frac{\partial V}{\partial z}\Bigr|_{\braket{z}}$ and  $\frac{d \braket{z}}{dt} = \frac{\braket{p}}{m}$, which allows us to write :
\begin{equation}
V_{0}=V(\braket{z}) + \frac{d}{dt}\,\left(\braket{z}\,\braket{p}\right) -\frac{ \braket{p}^2}{m}+\frac{1}{2}\frac{\partial^2 V}{\partial z^2}\Bigr|_{\braket{z}}\braket{z}^2
\end{equation}
hence we get 
\begin{align}
\frac{d \mathcal{I}\text{m}C}{dt}&=\frac{\hbar}{2\,m}\,\left[\langle z \rangle^2\left[(\mathcal{R}\text{e}A)^2-(\mathcal{I}\text{m}A)^2\right]-2\,\frac{\langle p\rangle\,\langle z\rangle}{\hbar}\,\mathcal{I}\text{m}A - \mathcal{R}\text{e}A\right] \nonumber\\ 
&-\frac{1}{2\hbar}\,\frac{\partial^2 V}{\partial z^2}\Bigr|_{\braket{z}}\braket{z}^2-\frac{1}{\hbar}\, \frac{d}{dt}\,\left(\braket{z}\,\braket{p}\right) +\frac{1}{\hbar}\,\left(\frac{\langle p\rangle^2}{2m}-V(\braket{z})\right)
\end{align}
\\
Now using the imaginary part of $A$ 
\begin{align}
\frac{d \mathcal{I}\text{m}A}{dt}&=-{\hbar\over m}((\mathcal{R}\text{e}A)^2- (\mathcal{I}\text{m}A)^2)+2\,\frac{V_{2}}{\hbar} \quad\text{with} &V_{2}=\frac{1}{2}\,\frac{\partial^2 V}{\partial z^2}\Bigr|_{\braket{z}}
\end{align}
we have then
 \begin{align}
\frac{d \mathcal{I}\text{m}C}{dt}&=-\frac{\hbar}{2\,m}\,\mathcal{R}\text{e}A-\frac{\langle z \rangle^2}{2}\,\left[\frac{d\mathcal{I}\text{m}A}{dt}   -\frac{1}{\hbar}\,\frac{\partial^2 V}{\partial z^2}\Bigr|_{\braket{z}} \right] -\frac{1}{m}\langle p\rangle\,\langle z\rangle\,\mathcal{I}\text{m}A\nonumber\\ 
&-\frac{1}{2\hbar}\,\frac{\partial^2 V}{\partial z^2}\Bigr|_{\braket{z}}\braket{z}^2-\frac{1}{\hbar}\, \frac{d}{dt}\,\left(\braket{z}\,\braket{p}\right) +\frac{1}{\hbar}\,\left(\frac{\langle p\rangle^2}{2m}-V(\braket{z})\right)
\end{align}
after some rearrangement, we get 
 \begin{equation}
\frac{d \mathcal{I}\text{m}C}{dt}=-\frac{\hbar}{2\,m}\,\mathcal{R}\text{e}A-\frac{1}{2}\,\frac{d}{dt}\left[\langle z\rangle^2\,\mathcal{I}\text{m}A\right]-\frac{1}{\hbar}\, \frac{d}{dt}\,\left(\braket{z}\,\braket{p}\right) +\frac{1}{\hbar}\,\left(\frac{\langle p\rangle^2}{2m}-V(\braket{z})\right)
\end{equation}
after integration, we get\footnote{The contribution $\frac{1}{\hbar}\int\,dt(-\frac{ \hbar^2}{2m}\,\mathcal{R}\text{e}A)$ can be considered as a gaussian correction to the usual, semi-classical contribution $\frac{1}{\hbar}\int\,dt \left(\frac{\langle  p\rangle^2}{2m}-V(\braket{z})\right)$. It is valid for any type of potential. As far as we know, this correction is never taken into account in atomic interferometry.}
\begin{equation}
\boxed{\mathcal{I}\text{m}C(t)=-\frac{1}{2}\langle  z\rangle^2\,\mathcal{I}\text{m}A-\frac{1}{\hbar}\braket{z}\braket{p}+\frac{1}{\hbar}\int\,dt \left(\frac{\langle  p\rangle^2}{2m}-V(\braket{z})-\frac{ \hbar^2}{2m}\,\mathcal{R}\text{e}A\right).}\label{great}
\end{equation}
Henceforth, using 
\begin{equation}
\mathcal{I}\text{m}A=-\frac{1}{\hbar\mathcal{Q}}\left[\mathcal{Q}\mathcal{P}-\frac{\hbar^2}{4}\right]^\frac{1}{2}
\end{equation}
we get
\begin{equation}\label{imcpmt}
\boxed{\mathcal{I}\text{m}C(t)=\frac{1}{2}\,\frac{\langle  z\rangle^2}{\hbar\,\mathcal{Q}}\left[\mathcal{Q}\mathcal{P}-\frac{\hbar^2}{4}\right]^\frac{1}{2}-\frac{1}{\hbar}\braket{z}\braket{p}+\frac{1}{\hbar}\int\,dt \left(\frac{\langle  p\rangle^2}{2m}-V(\braket{z})-\frac{ \hbar^2}{2m}\,\mathcal{R}\text{e}A\right)}
\end{equation}
with \begin{align}
\mathcal{Q}&=\langle z^2\rangle -\langle z\rangle ^2=\frac{1}{2\mathcal{R}\text{e}A}&\mathcal{P}&=\langle p^2\rangle -\langle p\rangle ^2=\frac{\hbar^2}{2}\,\frac{\vert A\,\vert^2}{\mathcal{R}\text{e}A}
\end{align}

\subsection{{\bf Classical action and quantum contributions to the phase shifts in the Humpty-Dumpty Stern Gerlach experiment}}
 Coming back to the Humpty-Dumpty experiment and collecting previous results from section \ref{temporal}, the Hamiltonian, inside the device, can be written in the following form:
\begin{equation}
\mathcal{H}_z=-\frac{\hbar^2}{2\,m}\,\frac{\partial^2}{\partial z^2}+\lambda\,\frac{g\mu_B}{2}\left(B_0-B'_0\,z\right)\otimes\sigma_z+V_\pm^G(z,t)
\end{equation}
with
\begin{align}
V_{\pm}^G(z,t)&=\underbrace{\nu_\pm^2\,\left[\frac{m}{2}\,\omega_{s}^2\left(z-\langle  z \rangle_\pm\right)^2+\frac{m\omega_{s}^2}{2}\,\mathcal{Q}_\pm(t)-\frac{6}{5}\,\frac{G\,m^2}{R}\right]}_\textrm{\small Self-interaction $\ket{\pm}\rightarrow \ket{\pm}$}-\underbrace{(1-\nu_\pm^2)\,\frac{Gm^2}{\vert \langle  z\rangle_+-\langle  z\rangle_-\vert}}_\textrm{\small Self-interaction $\ket{\mp}\rightarrow \ket{\pm}$}\label{temporalbis}
\end{align}
and with
\begin{align}
\nu_\pm&=\left\{ \begin{array}{rcl}
\small &1&\quad\text{if}\quad \vert \langle z \rangle_+ -\langle z \rangle_-\vert \leq 2R \\
&\vert \beta_\pm\vert&\quad\text{otherwise.}
\end{array}\right.&
\quad\text{and}\quad\lambda&=\left\{ \begin{array}{rcl}
\small &1&\quad\text{if}\quad 0\leq t\leq T_1 \quad \text{or}\quad T_4\leq t\leq T_5  \\
&0&\quad\text{if}\quad T_2\leq t\leq T_3\\
&-1&\quad\text{if}\quad T_1\leq t\leq T_2 \quad\text{or}\quad T_3\leq t\leq T_4
\end{array}\right.
\nonumber
\end{align}
Hence using the expression of the phase shifts for the spin up and spin down packets derived  before (\ref{c2sg}), we get:

{
\begin{equation}\label{c2sgnew}
\frac{d \mathcal{I}\text{m}C_\pm(t)}{dt}=\frac{\hbar}{2\,m}\,\left[\mathcal{R}\text{e}B_\pm^2-\mathcal{I}\text{m}B_\pm^2-\mathcal{R}\text{e}A_\pm\right]-\,\frac{V_{\pm}^G(t) \pm\lambda\frac{g\mu_B}{2}\,B_0}{\hbar}
\end{equation}
}
\tom{where $V_{\pm}^G(t)=V_{\pm}^G(<z_{\pm}>,t)$, and $V_{\pm}^G$ is defined at the level of equation (\ref{vgpm}/\ref{temporalbis}).}

After integration, we are free, making use of equation (\ref{imcpmt}),  to express the phase shift as the sum of a classical and a quantum contribution:
\begin{equation}
\mathcal{I}\text{m}C_\pm(t)=\underbrace{-\frac{\langle  z\rangle_\pm\langle  p\rangle_\pm}{\hbar}+\frac{1}{\hbar}\,\mathcal{S}_{Cl}}_\textrm{\small Classical contributions}+\underbrace{\frac{\langle  z \rangle_\pm^2}{2}\,\frac{1}{\hbar\mathcal{Q}_\pm}\left[\mathcal{Q}_\pm\mathcal{P}_\pm-\frac{\hbar^2}{4}\right]^\frac{1}{2}-\frac{1}{\hbar}\int\,dt~\mathcal{F}_{Q,\pm}}_\textrm{\small Quantum contributions}\label{small}
\end{equation}
where we defined the classical action $\mathcal{S}_{Cl}$ through
 \begin{align}
\mathcal{S}_{Cl,\pm}&=\int dt~\left[\frac{\langle  p\rangle_\pm^2}{2m}-V^{ext}(\langle  z \rangle_\pm)\right]\\\nonumber \end{align}
where $V^{ext}$ represents the magnetic potential $\lambda\,\frac{g\mu_B}{2}\left(B_0-B'_0\,z\right)\otimes\sigma_z$ while the contributions  of the self-gravitational potential $V_\pm^G(z,t)
$ have been included in the quantum contributions $\mathcal{F}_{Q,\pm}$:
\begin{align}
\mathcal{F}_{Q,\pm}&=\frac{\hbar^2}{4m\mathcal{Q}_\pm}+\frac{m\omega_{s}^2}{2}\,\mathcal{Q}_\pm(t)\,\nu_\pm^2-\frac{6}{5}\,\frac{G\,m^2}{R}\,\nu_\pm^2-(1-\nu_\pm^2)\,\frac{Gm^2}{d},\label{diantre}
\end{align}
with

\begin{align}
\mathcal{Q}_\pm&=\langle z^2\rangle _\pm-\langle z\rangle _\pm^2=\frac{1}{2\mathcal{R}\text{e}A_\pm}&\mathcal{P}_\pm&=\langle p^2\rangle _\pm-\langle p\rangle _\pm^2=\frac{\hbar^2}{2}\,\frac{\vert A_\pm\,\vert^2}{\mathcal{R}\text{e}A_\pm}
\end{align}

The terms $-\frac{\langle  z\rangle_\pm\langle  p\rangle_\pm}{\hbar}$ and $\frac{\langle  z \rangle_\pm^2}{2}\,\frac{1}{\hbar\mathcal{Q}_\pm}\left[\mathcal{Q}_\pm\mathcal{P}_\pm-\frac{\hbar^2}{4}\right]^\frac{1}{2}$ in equation ($\ref{small}$) are irrelevant regarding the phase shift because they cancel out during the recombination process (at the end of which $\langle  z \rangle_\pm=0$).
Knowing $\langle  z\rangle_\pm$, $\langle  p\rangle_\pm$ and $A_\pm(t)$ allows us to solve this equation and to deduce the phase difference between the two quantum paths $\ket{+}$ and $\ket{-}$. We derived in a previous section the expressions of $\langle  z\rangle_\pm$ and $\langle  p\rangle_\pm$, using Ehrenfest's theorem; we also derived an analytic expression of the function $A_\pm(t)$ (see appendix). In conclusion, up to an integration over time, all contributions to the phase shift are known in terms of analytic functions, which considerably enhances the precision of the numerical simulations.
\section{{\bf Numerical simulations}\label{numer}}

\begin{figure}\begin{center}\includegraphics[scale=0.53]{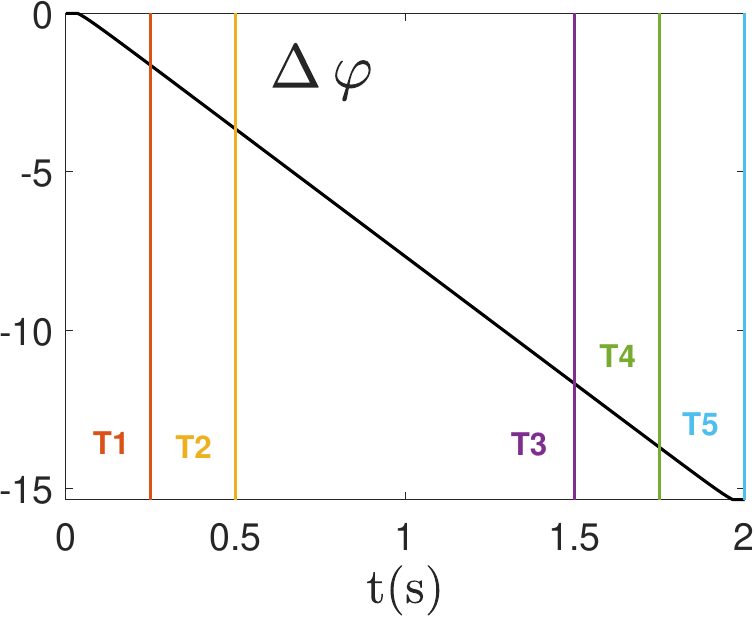}\caption{Plot of the phase shift accumulated in the evolution, with $T_3-T_2=1$ s.  We considered a mesoscopic sphere of radius $R=1\cdot 10^{-6}$ m with a mass $m=5,5.10^{-15}$ kg so that $\omega_s=\sqrt{\frac{G\,m}{R^3}}\sim 6.10^{-4}$ Hz. Here we also used $\vert \beta_+\vert=\frac{1}{\sqrt{3}}$ and $\vert \beta_-\vert=\sqrt{\frac{2}{3}}$ and $\sqrt{\mathcal{Q}_0}=10^{-9}$ m.}\label{phisg2}\end{center}\end{figure}

Let us define the phase shift as
\begin{equation}\label{shiftphi}
\Delta\varphi(t)=c_+(t)-c_-(t).
\end{equation}
To estimate the phase shift, we used the same parameters as in \cite{bosesg}. We consider thus a mesoscopic mass $m=5.5\cdot10^{-15}$ kg with radius $R=1\cdot 10^{-6}$ m and we used a field gradient $B_0'=10^{6}$ T.m$^{-1}$. Moreover, we considered $\vert \beta_+\vert =1/\sqrt{3}$ and an initial spread in position  $\sqrt \mathcal{Q}_0=10^{-9}$ m (other initial spreads were also considered in the supplementary material).
\moh{The expressions of the trajectories $\langle  z\rangle_\pm$ and the momentum $\langle  p\rangle_\pm$ of each wave packet are plotted in figure \ref{sghp}.}  When the two wave packets $\psi_L$ and $\psi_R$ are recombined (using a magnetic field oriented in the opposite direction), the state becomes:
\\
\begin{equation}
\Psi(z,t)=\left[\beta_+\ket{+}+\beta_- e^{-i\,\Delta\varphi (T_5)}\,\ket{-}\right]\,\psi(z,t)\label{deltaphi5}
\end{equation}
\\
where $T_5$ is the total time of the experiment. In figure \ref{phisg2} we illustrate the phase shift $\Delta\varphi $ accumulated during this evolution in function of time. \moh{We chose $T_5= 2$ s (and $T_3-T_2=$ 1 s), which corresponds to the evolution plotted in figure \ref{sghp})}.

Actually, if we have a strict equality $\vert \beta_+\vert=\vert \beta_-\vert=1/\sqrt{2}$ we find at the end of the Humpty-Dumpty experiment that the phase shift is zero, as it must be due to symmetry.  Because of symmetries, there is also no classical contribution to the final value of the phase shift. Therefore we only plot the contributions made by the quantum term $\mathcal{F}_{Q,\pm}$ present in equation \eqref{diantre}. Although this phase shift consists of several non-trivial quantum contributions, we can estimate it as follows. If we naively only take account of the contribution $\frac{6}{5}\,\frac{G\,m^2}{\hbar\,R}\,(T_5-2\,T_s)\,\left(\vert \beta_+\vert^2- \vert \beta_-\vert ^2\right)$ we expect to find, for $T_5-2\,T_s=1.93$ s, a phase shift of the order of $-15.59$ which is almost the exact value $\Delta\varphi \sim-15.33$.
 In the regime of parameters that we considered, it can be shown (see appendix, sections \ref{sgap1} and \ref{nuclear}) that in good approximation $A_+(t)\sim A_-(t)\sim A_0(t)$ where $A_0(t)$ corresponds to a freely evolving gaussian wave packet. This is so because either the pulsation of the (comoving) harmonic potential $\omega_{s}$ is very small or, when it is large due to nucleic contributions (see sections \ref{bla} and \ref{nuclear}), it is large only during very short times. This explains why the naive estimate involving only the contribution $\frac{6}{5}\,\frac{G\,m^2}{\hbar\,R}\,(T_5-2\,T_s)\,\left(\vert \beta_+\vert^2- \vert \beta_-\vert ^2\right)$ dominates other contributions, as can be seen from figures \ref{conte} and \ref{contf} where all contributions (which are also detailed in appendix) appearing at the level of equation (\ref{diantre}) are plotted separately, for two distinct initial conditions, differing by the initial size of the wave packet.
 
 \begin{figure}[!h]
\begin{center}
	\includegraphics[scale=0.67]{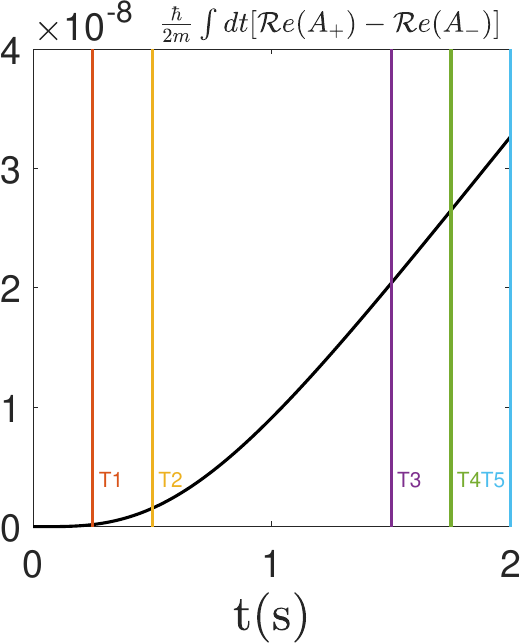}
	\includegraphics[scale=0.67]{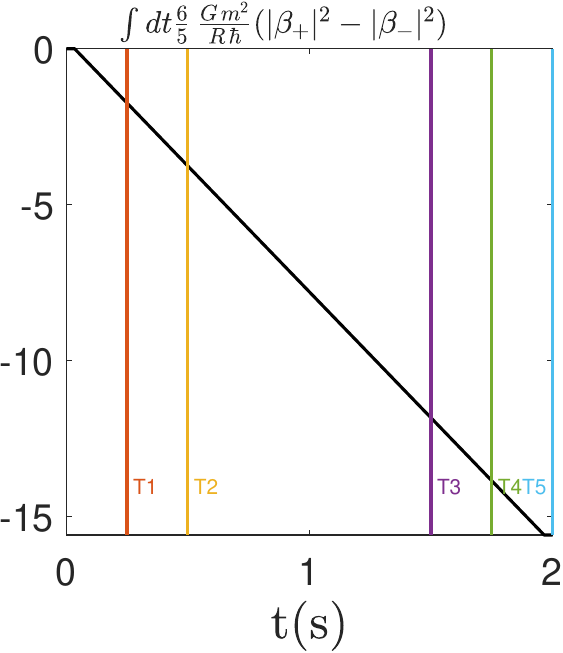}
	\includegraphics[scale=0.67]{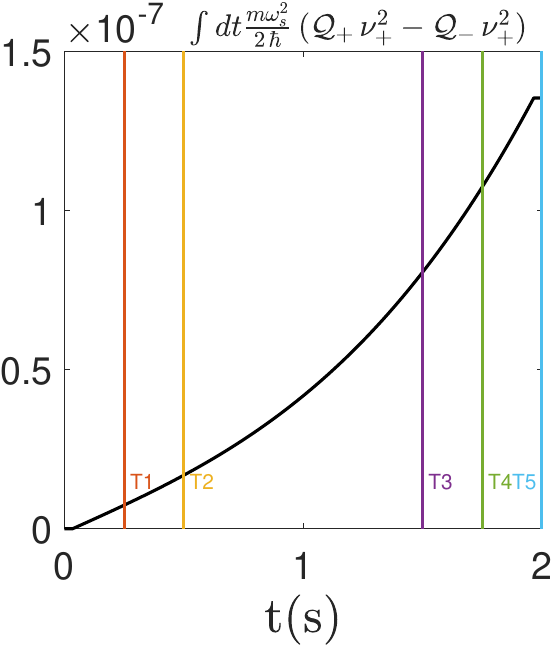}
	\includegraphics[scale=0.67]{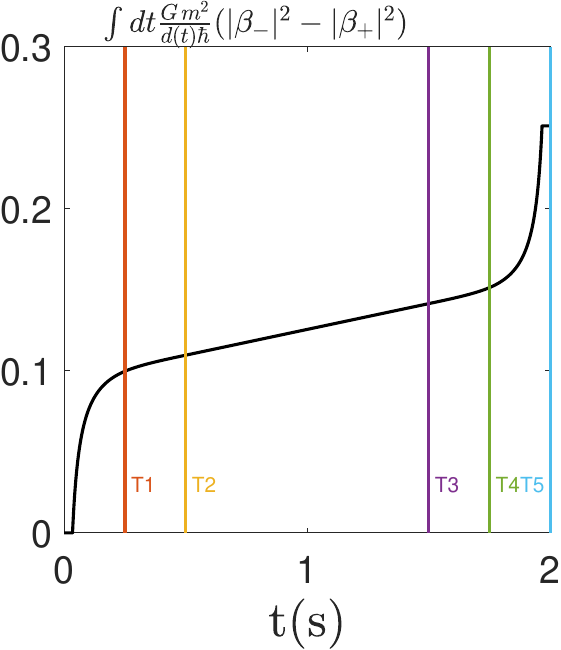}
	\caption{Plots of individual contributions of each term in the expression (\ref{diantre})  of the total phase shift. We chose $\sqrt{\mathcal{Q}_0}=10^{-10}$ m, $m=5,5.10^{-15}$ kg, $\omega_s=\sqrt{\frac{G\,m}{R^3}}\sim 6.10^{-4}$ Hz, $\vert \beta_+\vert=\frac{1}{\sqrt{3}}$ and $\vert \beta_-\vert=\sqrt{\frac{2}{3}}$.}\label{conte}
\end{center}
\end{figure}
\begin{figure}
\begin{center}
	\includegraphics[scale=0.67]{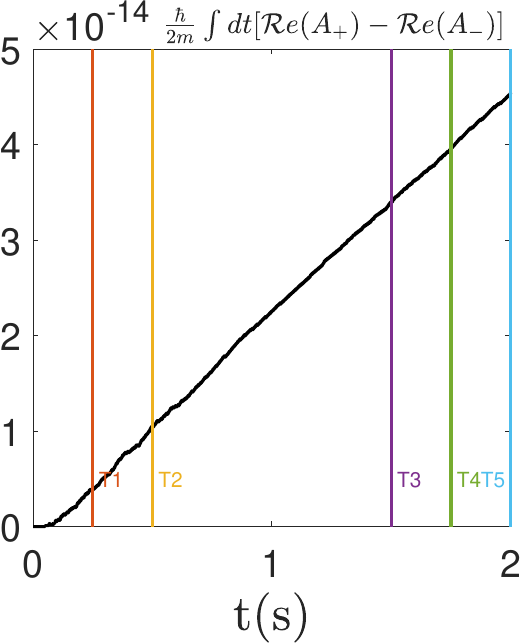}
	\includegraphics[scale=0.67]{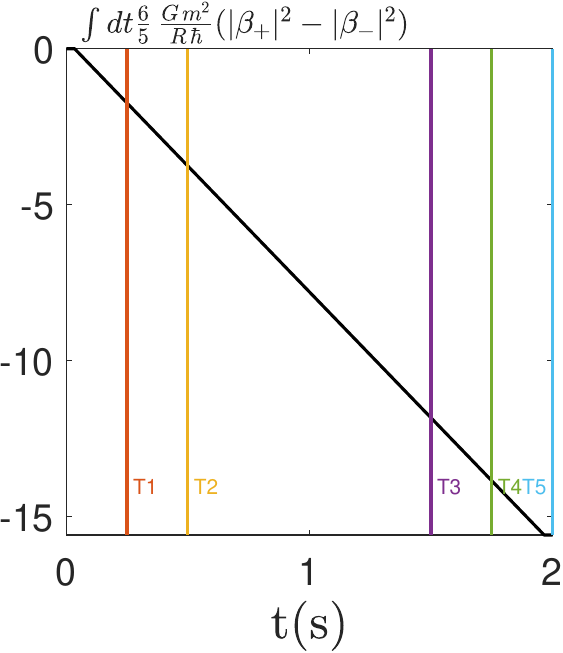}
	\includegraphics[scale=0.67]{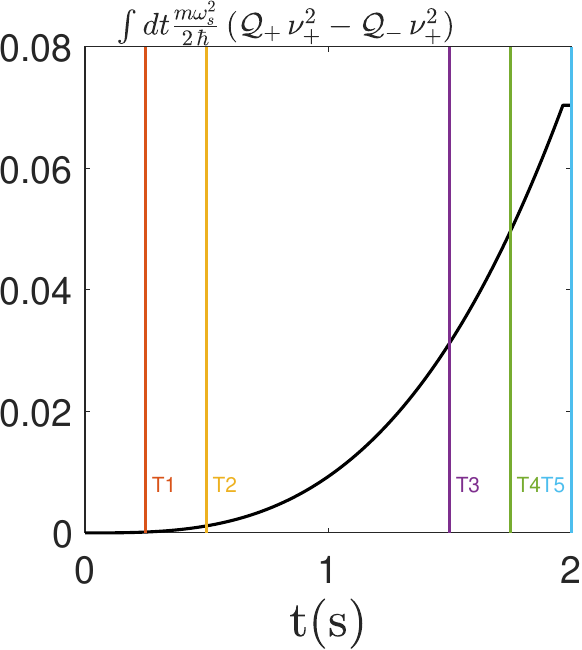}
	\includegraphics[scale=0.67]{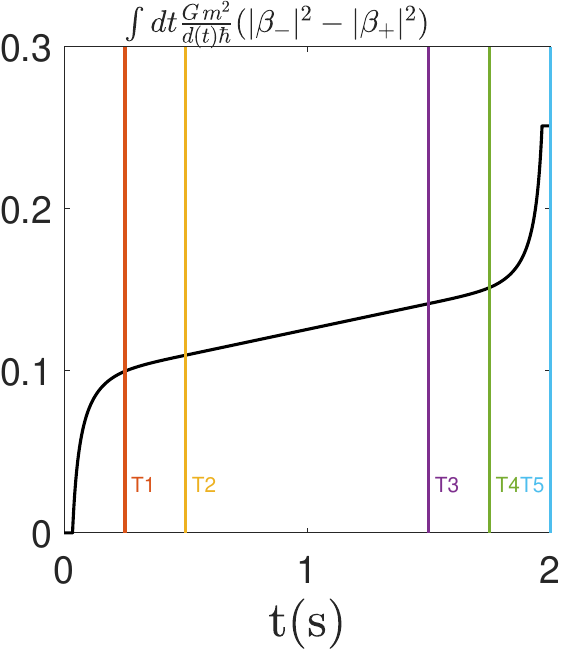}
	\caption{Plots of numerical simulations of the individual contributions of each term in the expression (\ref{diantre})  of the total phase shift. We chose $\sqrt{\mathcal{Q}_0}=10^{-13}$ m, $m=5,5.10^{-15}$ kg, $\omega_s=\sqrt{\frac{G\,m}{R^3}}\sim 6.10^{-4}$ Hz, $\vert \beta_+\vert=\frac{1}{\sqrt{3}}$ and $\vert \beta_-\vert=\sqrt{\frac{2}{3}}$.}\label{contf}
\end{center}
\end{figure}

 \tom{ It is worth noting that, in our approach, no spin decoherence appears after recombination, in good approximation, because internal (spin) and external (spatial) degrees of freedom are not entangled in the scenario described here. This is due to the fact that the wave packets nearly perfectly overlap at the end of the recombination process (along $X$, $Y$ and $Z$ as well), not only because their centers and wave vectors are assumed to coincide perfectly, but also because they have the same shape ($A_+(t)\sim A_-(t)\sim A_0(t)$), as explained in appendix, sections \ref{sgap1} and \ref{nuclear}.  In other words, there is no which-path information induced by the dynamics. This high degree of coherence contrasts with more realistic experimental situations described, for instance, in Refs. \cite{robert2001,julius,folman,henkel22}. }
  

 \section{Discussions\label{disc}}
  \subsection{Some orders of magnitude.}\label{bipbip}
\begin{figure}[h]
\begin{center}
	\includegraphics[scale=0.73]{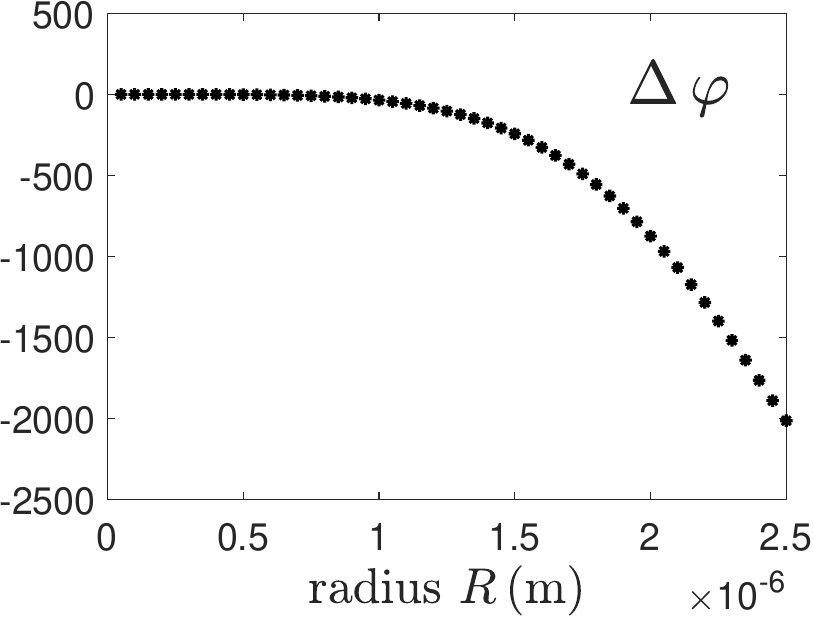}
	\caption{Plot of the accumulated phase shift in function of the radius $R$ of the nanosphere. We chose the same magnetic fields and time-steps as before, and $\vert \beta_+\vert=\frac{1}{\sqrt{3}}$ and $\vert \beta_-\vert=\sqrt{\frac{2}{3}}$ and an initial spread $\sqrt{\mathcal{Q}_0}=10^{-9}$ m, as in figure \ref{phisg2}.}\label{phiR}
\end{center}
\end{figure}

 \moh{
The main progress in our proposal, as well as in the proposals of references \cite{bosesg,vedral}, is that they are based on an interferometric effect. Other proposals in the past \cite{CDW,CDWPRA,Giulini2011,Meter2011} aimed at measuring the inhibition of the spread of a wave packet due to self-gravity concluded that free-fall times of the order of $10^4$ seconds are required in order to get an observable effect. A gaussian wave packet gets indeed frozen by self-gravity if its size is of the order of the Lieb radius $\hbar^2/Gm^3$. At twice the normal density, the mesoscopic transition occurs when $R=\hbar^2/Gm^3\approx 10^{-7}$ m which corresponds to a mass of the order of $10^{-17}$ kg.  The corresponding spread in absence of self-gravity would then be equal to $\mathcal{Q}_0\left[1+ \frac{\hbar ^2\,t^2}{4 m^2 \mathcal{Q}_0^2}\right]$. In order to be able to differentiate both spreads in this regime with an accuracy of one micron, we must impose $\mathcal{Q}_0=R^2$ and $\frac{\hbar ^2\,t^2}{4 m^2 \mathcal{Q}_0^2}\approx100$ which requires \cite{CDW} free fall times of the order of 10$^4$ s \cite{CDW,Giulini2011,Meter2011}, which are impossible to realize on earth and require the use of a satellite.
\\
In an interferometer, the situation is totally different, it is the gravitational energy (of the order of) $(Gm^2/d)$ which dominates the contributions to the dephasing. If we impose, for instance, $T_5=1$ s and $d=100~\mu \text{m}$ as in \cite{bosesg} and that we require the dephasing (which is of the order of $(Gm^2T_5/d)/\hbar$) to be of the order of unity, this fixes the value of the \moh{mass ($\approx 8,7.10^{-15}$ kg in this case which corresponds with a density $\rho\approx1300 \text{Kg.m}^{-3}$ to a radius a bit larger than $1~\mu \text{m}$).} This is illustrated in figure \ref{phiR} where we plot the phase shift for different values or the radius $R$ of the nanosphere, i.e. different values of the mass; it is clear that a minimal \moh{radius of the order of $1~\mu \text{m}$. is required to see the effect.}}

\subsection{One versus Two Humpty-Dumpty devices.}
\label{bopbop}
At this level, we may relax the original assumption according to which it was necessary to prepare the initial state in the ground state of the optical trap from which it is released before entering the S-G apparatus. The reason that we advocated for doing so was that we needed a pure gaussian state to begin with. Retrospectively we see that, as it is the additive constant $\frac{6}{5}\,\frac{G\,m^2}{\hbar\,R}$ in the self-interaction that mainly contributes to the dephasing, a similar dephasing is expected to occur even when the center of mass degrees of freedom are initially prepared in a thermal state. \tom{If the experiment reveals that no dephasing is present, this would mean that there is no gravitational self-interaction (\ref{classgrav}) in nature; then, a double Humpty-Dumpty experiment as already proposed in \cite{bosesg,vedral} would make it possible to measure, if it exists, the entangling power of gravity and to test the validity of the electro-magnetic analogy, opening the door to various models among which quantum gravity models \cite{hua2014} but also possibly other models inspired by the electro-magnetic analogy of section \ref{analogy}.} The double Humpty-Dumpty experiment is more difficult to realize than our proposal, not only because two interferometers must be realized in parallel. In order to minimize Casimir-Polders interaction, the authors of  \cite{bosesg} must impose that the distance between the objects is de facto quite larger than their size (of the order of 100 times larger), so that free fall times of the order of at least 2.5 seconds are necessary (working in the same conditions) to see a dephasing of the order of one radian. 

When only one object is present, as in our case, the Casimir-Polders force is equal to zero because there is no electromagnetic self-interaction in nature, as we explained at the beginning of this paper. All that is necessary in our case is to nullify the spatial overlap of the up and down wave packets. Our experimental proposal would thus still be feasible with quite less intense magnetic fields and/or quite a shorter experiment time than those required for realizing the double Humpty-Dumpty proposals  \cite{bosesg,vedral}. For instance, if we impose that the up and down components fall side by side, at a distance $d_\pm=\frac{g\mu_B}{m}\,B'_0\,\widetilde{T_1}^2$ of the order of $2\,R=2.10^{-6}$ m, we may let act the magnets at the same intensity as before but with a time almost ten times shorter, $\widetilde{T_1}\sim T_1/10=0.025$ s; we may also diminish all the time intervals $T_i$ and then the time of the experiment in the same ratio: $\widetilde{T_5}\sim 0.2$ s. After recombination the dephasing will be of the order of $\left[(-\frac{6}{5}\,\frac{G\,m^2}{\hbar\,R}+\frac{Gm^2}{2\,R\,\hbar})\cdot\left(\widetilde{T_5}-2\,T_s\right)\cdot\left(\vert \beta_+\vert^2 -\vert\beta_- \vert^2\right)\right]\approx -1.1+0.4=-0.7$. Compared to the double Humpty-Dumpty experiment where comparable dephasings require a free fall of 2,5 s (thus a vertical distance larger than 30 m, only reachable in a free fall tower), our proposal requires a distance of $\frac{g\,\widetilde{T_5}^2}{2}=$20 cm which can be done on a lab. table.\\

 \subsection{Why in good approximation, the internal density of the object (rigid sphere) can be considered to be constant (homogeneous).}\label{bla}
\tom{ If the size of the wave function of the center of mass of the freely falling object is comparable to the spread of the nucleons inside its crystalline structure, it is not appropriate to treat this object as a sphere of constant density. This is so because the self-interaction of the nucleons must be taken into account. This property boosts the self-interaction and in particular, it boosts the harmonic self-trapping by increasing the value of the parameter $\omega_s$ introduced at the level of equation (\ref{fullpotNS2}) in a ratio of the order of one thousand as is shown in section \ref{nuclear}. This feature has been exploited in the experimental proposal of Yang {\it et al.}\cite{chen}  aimed, like ours, at revealing the existence of a self-interaction {\it \`a la} S-N. Now, in their proposal, Yang {\it et al.} assumed that the object was placed in a harmonic trap. Here, we consider a freely falling object, which makes a huge difference. Indeed, if in a trap, it is possible to guarantee that at all times, the size of the wave packet associated with the center of mass remains smaller than the size of the nucleons (which is of the order of $10^{-12}$ m \cite{chen}), this is an impossible task in the case of freely falling objects, excepted for unreasonable parameter choices, in virtue of Heisenberg uncertainty principle. Consider, for instance, an initial spread in position $\delta x_0$, at time  $T_2$, smaller than or equal to $10^{-12}$ m and a mass $m$ of at most $10^{-14}$ kg, then, the spread in velocity which is of the order of ${\hbar\over m\cdot \delta x_0}$ will certainly be larger than or equal to $10^{-8}$ m/s. This means that after a time of the order of $10^{-2}$ s, the spread in position will be a hundred times larger than the size of the nucleons, in which case it is consistent to neglect the nucleic contributions to $\omega_s$. Now, in order to put into evidence gravitational effects, we need $T_3-T_2$ to be at least of the order of 0,1 s as discussed in section \ref{bipbip}, which is longer than $10^{-2}$ s so that we expect a small netto benefit after preparing the initial wave packet in a region smaller than the size of the nucleon, even if we take into account the boost of the value of $\omega_s$. This is corroborated by the estimates presented in appendix (section \ref{nuclear}), where we show that the relevant, dimensionless, parameter measuring the intensity of the nuclear corrections is shown to be $\omega_s\cdot t$. For a time shorter than $10^{-2}$ s,  a boosted value of $\omega_s$ of 1 Hz, $\omega_s\cdot t$ remains smaller than 10$^{-2}$, which is small.  To conclude this section, taking the self-interaction of the nucleons into account either leads to small and nearly unobservable phase shifts or implies unreasonable parameter choices.  }
\tom{ It is \green{thus} wrong to neglect the dominating contribution to the phase shift (which is of the order of $\frac{6}{5}\,\frac{G\,m^2}{\hbar\,R}\,(T_5-2\,T_s)\,\left(\vert \beta_+\vert^2- \vert \beta_-\vert ^2\right)$ as shown by us in section \ref{numer}), contrary to what is claimed in reference \cite{inhib} concerning the constant contribution of the gravitational self-energy to the potential where one can read the following: \it{...the constant contribution (...) appears nowhere, only its derivatives enter into any observable quantity....} }\green{This ill-founded hypothesis is equivalent in our eyes with throwing the baby with the water of the bath.}

  \subsection{Developing the electromagnetic analogy.}\label{analogy}
  
  \tom{Compared to all other physical theories, QED has been confirmed with unprecedented accuracy, and it is obviously a vivid source of inspiration for quantum gravity\cite{hua2014}. Moreover, even in the non-relativistic limit, the incorporation of the Coulomb interaction into the non-relativistic Schr\"odinger equation in two simple cases, such as hydrogen and helium atom, leads to accurate predictions confirmed by experiments so that the corresponding models can serve as a source of inspiration for a toy-model of quantum gravity in the non-relativistic limit, to be contrasted with the S-N interaction (\ref{multibody}) considered here. }
 
  \tom{  Let us consider first the case of the hydrogen atom.
Replacing the mass $m$ by the charge of the electron, Newton's constant $G$ by the Coulomb constant, and self-attraction by self-repulsion, in equation (\ref{NS}), we obtain the Wigner-Poisson equation, which has been successfully implemented in plasma physics or solid state physics in order to mimic repulsive Coulomb self-interaction between many electrons in the mean-field (Hartree) regime. In the single-particle case, however, the Wigner-Poisson equation is clearly not relevant and ruled out by facts, among others, because  if we would apply it to quantize electronic energy levels in the hydrogen atom it would drastically modify Bohr's spectrum, strongly contradicting accurate spectroscopic data accumulated since the 19th century by Rydberg and others  \cite{CDW,hua2014}. This simple example rules out semi-classical models such as the droplet model originally associated by Schr\"odinger to the electron \cite{zeit}, where $|\Psi|^2$ would represent a density of stuff (here charge). It also helps to understand why quantum gravity and semi-classical theories of gravity {\it \`a la} S-N are very likely to lead to incompatible predictions as is shown in the present paper. Among others, a semi-classical treatment of charge/mass leads to the appearance of self-interaction \cite{sebens}, which is not expected to occur in ``QED inspired'' approaches like quantum gravity \cite{hua2014}. }

\tom{It is also instructive to develop the electro-magnetic analogy to the case of the helium atom where two objects (here two electrons denoted $A$ and $B$) are present (the proton being associated to the external Coulomb potential). If we pursue the analogy with electro-magnetism, we find that a ``quantum gravitational analog'' of the two-electron interaction term would be, after replacing charges by masses and Coulomb constant by minus Newton constant,} \green{the following { linear} potential:}
\begin{equation}
-\int~d^3x_A\,\int~d^3x_B'~\vert\Psi(t,{\bf x}_A,{\bf x'}_B)\vert^2\,\frac{G\,m^2}{\vert{\bf x}_A-{\bf x'}_B\vert},\label{quantgrav}
\end{equation}where $|\Psi(t,{\bf x}_A,{\bf x'}_B)\vert^2=\braket{\Psi\vert\Psi}$ with 

\begin{align}\ket{\Psi}=\alpha (t)\psi_{++}((t,{\bf x}_A,{\bf x'}_B)\ket{+^A+^B}+\beta (t)\psi_{+-}((t,{\bf x}_A,{\bf x'}_B)\ket{+^A-^B}\nonumber \\
+\gamma (t)\psi_{-+}((t,{\bf x}_A,{\bf x'}_B)\ket{-^A+^B}+\delta (t)\psi_{--}((t,{\bf x}_A,{\bf x'}_B)\ket{-^A-^B}\end{align}
\tom{If each packet $\psi_{i,j}$ (with $i,j=\pm $) is localized in a tiny region of space, and so the four spin components are distant from each other, the potential of interaction is essentially the Newton potential, and the electro-magnetic analogy leads to the same predictions regarding the entangling power of the interaction as those made in References \cite{bosesg,vedral}.} \green{Actually, the effective interaction encapsulated at the level of equation (\ref{quantgrav}) has been derived rigorously combining perturbed linearized general relativity and quantum field theory while its electro-magnetic analog can also be derived rigorously in a QED approach, as discussed in Ref. \cite{hua2014}.}


\tom{In order to estimate the entangling power of the evolution (\ref{quantgrav}) it suffices to note that if at time $t=0$ the system is prepared in a spin state 
$\alpha \ket{+^A+^B}+\beta \ket{+^A-^B}+\gamma \ket{-^A+^B}+\delta \ket{-^A-^B}$, where the various spin components are localized in distant regions of space as in figure \ref{fig1}, and that we denote $d_{+^A+^B}$, $d_{+^A-^B}$, $d_{-^A+^B}$ and $d_{-^A-^B}$ the average distances between these components, then after a free fall of duration $T$ the spin state becomes}
\begin{align}
    \alpha e^{-i{-Gm_A\cdot m_B\cdot T\over \hbar d_{+^A+^B}}} \ket{+^A+^B} +& \beta e^{-i{-Gm_A\cdot m_B\cdot T\over \hbar d_{+^A-^B}}}\ket{+^A-^B} \nonumber \\ +&\gamma e^{-i{-Gm_A\cdot m_B\cdot T\over \hbar d_{-^A+^B}}}\ket{-^A+^B}+\delta e^{-i{-Gm_A\cdot m_B\cdot T\over \hbar d_{-^A-^B}}}\ket{-^A-^B}
\end{align}
\tom{In particular, if, at time $t=0$, the two objects are disentangled, which means that $\alpha\delta=\beta\gamma$, then, at time $t$, we get}
\begin{align*}
    \alpha(t)\delta(t)=e^{-i(-{Gm_A\cdot m_B\cdot T\over \hbar}\cdot ({1\over d_{+^A+^B}}+{1\over d_{-^A-^B}})} \alpha\delta \quad\text{and}\quad \beta(t)\gamma(t)=e^{-i(-{Gm_A\cdot m_B\cdot T\over \hbar}\cdot ({1\over d_{+^A-^B}}+{1\over d_{-^A+^B}})}\beta\gamma
\end{align*}
\tom{ and, so that generically systems $A$ and $B$ will be entangled (unless, by chance ${1\over d_{+^A+^B}}+{1\over d_{-^A-^B}}={1\over d_{+^A-^B}}+{1\over d_{-^A+^B}}$), in full agreement with the predictions of references \cite{bosesg,vedral}.} \\
\tom{As we shall now show, this intuitive picture must be abandoned in the semi-classical approach, and leads to predictions concerning the entangling power of the interaction which differ from those obtained in the quantum gravity approach. }
 

\section{Entangling power of semi-classical interaction {\it \`a la} Schr\"odinger-Newton.\label{zero}}
\tom{We estimated in previous sections the spin dephasing resulting from self-interactions. Here we shall also consider the dephasing resulting from distant, Newton-like interactions between two objects freely falling along parallel trajectories as plotted in figure \ref{fig1} and discuss the entanglement generated during such interaction. Let us denote these objects $A$ and $B$ as in the previous section. }
\tom{In virtue of equations (\ref{previouseqn1},\ref{previouseqn2}), we find that, during the free fall, the effective potential felt by the $i$ ($i=\pm$) spin component of the object $B$ due to the gravitational interaction with the $\pm$ spin component of the  object $A$ is equal in good approximation to ${-Gm_A\cdot m_B|\beta_\pm^A|^2\over d_{\pm^A,i^B}}$ where\footnote{The shift is proportional to $m\cdot |\beta_\pm^A|^2 $ and scales thus quadratically with $|\beta_\pm^A|$ (in agreement with our discussion of section \ref{previous}), it scales neither quartically (in $|\beta_\pm^A|^2|\beta_\pm^B|^2$) as expected in a classical model of interaction nor as $m\cdot |\beta_\pm^A|^0 $, as predicted in the framework of the electro-magnetic analogy developed in section  \ref{analogy} (and as assumed in Ref.\cite{inhib}).} $|\beta_+^A|^2=|\alpha|^2+|\beta|^2$ and $|\beta_-^A|^2=|\gamma|^2+|\delta|^2.$ Similarly, the effective potential felt by the $j$ ($j=\pm$) spin component of the object $A$ due to the gravitational interaction with the $\pm$ spin component of the  object $B$ is equal in good approximation to ${-Gm_A\cdot m_B|\beta_\pm^B|^2\over d_{j^A,\pm^B}}$ where $|\beta_+^B|^2=|\alpha|^2+|\gamma|^2$ and $|\beta_-^B|^2=|\beta|^2+|\delta|^2. $ Let us denote $\alpha \ket{+^A+^B}+\beta \ket{+^A-^B}+\gamma \ket{-^A+^B}+\delta \ket{-^A-^B}$ the spin state of the full system at time $t=0$. Then, after a free fall of duration $T$, the spin state becomes}
\begin{align*}
    \alpha e^{-i{-Gm_A\cdot m_B\cdot T\over \hbar}\cdot({|\beta_+^A|^2+|\beta_+^B|^2 \over d_{+^A+^B}}+{|\beta_-^B|^2 \over d_{+^A-^B}}+{|\beta_-^A|^2 \over d_{-^A+^B}})-i(\phi^{self}_{A+}+\phi^{self}_{B+})} \ket{+^A+^B}  \\
    +\beta  e^{-i{-Gm_A\cdot m_B\cdot T\over \hbar}\cdot({|\beta_+^A|^2+|\beta_-^B|^2 \over d_{+^A-^B}}+{|\beta_+^B|^2 \over d_{+^A+^B}}+{|\beta_-^A|^2 \over d_{-^A-^B}})-i(\phi^{self}_{A+}+\phi^{self}_{B-})} \ket{+^A-^B}  \\
    +\gamma e^{-i{-Gm_A\cdot m_B\cdot T\over \hbar}\cdot({|\beta_-^A|^2+|\beta_+^B|^2 \over d_{-^A+^B}}+{|\beta_-^B|^2 \over d_{-^A-^B}}+{|\beta_+^A|^2 \over d_{+^A+^B}})-i(\phi^{self}_{A-}+\phi^{self}_{B+})}\ket{-^A+^B} \\
    +\delta e^{-i{-Gm_A\cdot m_B\cdot T\over \hbar}\cdot({|\beta_-^A|^2+|\beta_-^B|^2 \over d_{-^A-^B}}+{|\beta_+^B|^2 \over d_{-^A+^B}}+{|\beta_+^A|^2 \over d_{+^A-^B}})-i(\phi^{self}_{A-}+\phi^{self}_{B-})}\ket{-^A-^B}
\end{align*}
\tom{ where $\phi^{self}_{A(B)\pm}=\mathcal{I}\text{m}.C_{A(B)\pm}$ where $C_\pm$ has been defined at the level of equation (\ref{gauss}) while $\mathcal{I}\text{m}.C_{A(B)\pm}$ has been shown to obey equation (\ref{imcpmt}).
It is easy to check that if at time $t=0$ $\alpha\gamma=\beta\gamma$ then $\alpha\gamma=\beta\gamma$ at all times and the entangling power of the semi-classical gravitational interaction is equal to zero (see section \ref{appendpower} in appendix for another, independent, derivation of this property).} The deep reason of this property is that the S-N potential due to the presence of, say, the $B$ system at the level of, say, the location of the spin up component of the $A$ system,  is the same, that we consider the $\ket{+^A+^B}$ state or the $\ket{+^A-^B}$ state. Obviously, this property is very general and we expect therefore that the entangling power of the S-N interaction will always be equal to zero, in all situations. \green{Incidentally, this result confirms the intuition proposed in reference \cite{bosesg,vedral}, according to which no classical interaction would make it possible to entangle the systems $A$ and $B$. }

\section{{\bf Conclusions} \label{conc}}
 \green{As we explained in the introduction, including a non-linearity at the level of Schr\"odinger equation marks a strong departure from other proposals aimed at incorporating the gravitational interaction in a quantum framework. One of the motivations for doing so is that it could contribute to solving the measurement problem \cite{Colin2017}, but other options remain open \cite{seven}. In reference \cite{hua2014}, for instance, a QFT approach based on perturbed linear gravity is shown to lead to a linear potential with no self-interaction, similar to the one described in section \ref{analogy}. General relativity being non-linear it is not clear whether its quantized version ought to be linear or not, and it could be that a satisfactory quantum description of gravity will impose to go beyond general relativity and quantum field theory. This explains why we focus here on a realizable experimental proposal: we are convinced that more experimental data about gravity in the quantum regime are needed if we wish to progress in this field \cite{seven}.  } 

\tom{This paper has been written in three steps. Preliminary results were presented for the first time in the Ph.D. thesis of one of us (M.H.) in 2019 \cite{PhD}. Thereafter we managed to estimate the various contributions contained in the final phase shift using both semi-analytical and numeric methods separately; among others, we estimated in section \ref{diantrebis} the gaussian correction (\ref{great}) to the semi-classical expression of the phase-shift commonly in use in atomic interferometry. These results led to a reprint in 2020 \cite{arxiv}. In the present paper we definitively clarified some misconceptions \cite{inhib} regarding the scaling of the self-interaction in function of the amplitudes $\beta_\pm$. This paved the way for an estimate of the entangling power of the S-N interaction which is rigorously shown to be null in section \ref{zero}. The present paper contains all these contributions. One of the reasons why it took such a long time to achieve this work is the covid crisis, and its unexpected overload of teaching duties in the case of one of us (T.D.).} 

Due to the non-linear nature of the S-N interaction, some approximations were necessary in order to be able to tackle the problem but all our approximations are rigorously derived and justified in a fully explicit and transparent fashion throughout the paper. For the rest, our derivations are exact and our results possess an analytical expression. We are thus confident that if one day the experiment described here gets realized, it will undoubtedly constitute a crucial experiment regarding the existence S-N interaction. \moh{As has been shown here, measuring the phase shift $\Delta\varphi (T_5)$ would  (see equation \eqref{deltaphi5}), indeed, enable us to establish the existence of a self-gravitational interaction {\it \`a la} S-N, which is the main result of our paper\footnote{Note that realizing a Stern-Gerlach Humpty-Dumpty experiment in the regime considered in the present paper would also make it possible to test other non-standard models like e.g. the de Broglie-Bohm interpretation \cite{xmas} and, possibly, a large range of models of spontaneous localization \cite{seven}}.} \tom{Nano-resonators, for which cooling of the center of mass degree of freedom and embedding of NV centers have already been successfully demonstrated in the past  \cite{arcizet2011,yuan2015} are promising candidates for testing our ideas, but the most promising candidate can be attributed without a doubt to Folman's team, because of its expertise in Stern-Gerlach experiments, and its explicit objective to realize a Humpty-Dumpty Stern-Gerlach interferometer with massive objects \cite{folman,folman21}.}

In any case, all these proposals constitute a breakthrough in the sense that they aim at measuring gravitational effects originating from delocalized objects in the mesoscopic regime. It is worth trying to realize them because they could provide the missing clues necessary for properly quantizing gravity. They could also contribute to clarifying the measurement problem and to discriminate between various interpretations of the quantum theory \cite{seven,xmas}. We are confident, in view of all the results presented in this paper, that if one day the experiment described here gets realized, it will constitute a crucial experiment regarding the existence of the S-N interaction. 
\section*{Acknowledgements}
The authors are pleased to acknowledge Alexandre Matzkin and Ralph Willox for fruitful discussions. We gratefully acknowledge funding and support from the John Templeton Foundation (grant 60230, Non-Linearity and Quantum Mechanics: Limits of the No-Signaling Condition, 2016–2019). MH acknowledges support from Aix-Marseille University for an ATER contract in 2019-2020 and support from the Okinawa Institute of Science and Technology Graduate University. 
\bibliographystyle{apsrev4-2}
\input{main.bbl}

\section{{\bf Appendix}}

\subsection{Estimate of $A_\pm(t)$}\label{sgap1}
Here the following equation for $A_\pm (t)$ is solved  
\begin{equation}
i\,\frac{d A_\pm(t)}{dt}=\frac{\hbar}{m}\,A_\pm(t)^2-2\,\frac{V_{2,\pm}(t)}{\hbar}\quad\text{where} \quad V_{2,\pm}(t)=\frac{m\,\omega_{s}^2}{2}\nu_\pm
\end{equation}
with
\begin{equation}
\nu_\pm=\left\{ \begin{array}{rcl}
\small &1&\quad\text{if}\quad \vert \langle  z \rangle_+ -\langle  z \rangle_-\vert \leq 2R \\
&\vert \beta_\pm\vert &\quad\text{otherwise.}
\end{array}\right.
\end{equation}
It is useful to use a set of dimensionless variables such that
\begin{equation}
X(t)=A_\pm(t)\,L^2\quad\text{and}\quad s=\omega_{s}\,t
\end{equation}
with
\begin{equation}
L^2=\frac{\hbar}{m\,\omega_{s}}
\end{equation}
Thus we get
\begin{equation}
 i\,\frac{d X}{ds}=X^2-\nu_\pm^2
\end{equation}
then
\begin{equation}
\frac{dX}{\left(X-\nu_\pm\right)\left(X+\nu_\pm\right)}=-i\, ds
\end{equation}
which can be put in the form 
\begin{equation}
\frac{dX}{2\,\nu_\pm}\,\left(\frac{1}{X-\nu_\pm}-\frac{1}{X+\nu_\pm}\right)=-i\,ds
\end{equation}
after integration we get
\begin{equation}
\ln\left(\frac{X-\nu_\pm}{X+\nu_\pm}\,\frac{X_0-\nu_\pm}{X_0+\nu_\pm}\right)=-2i\nu_\pm\,s
\end{equation}
\\
with $X_0=X(t=0)$ and $c_0=\frac{X_0-\nu_\pm}{X_0+\nu_\pm}$. Finally, it reads
\begin{equation}
\boxed{X(s)=\nu_\pm\,\frac{1+c_0\,e^{-2i\nu_\pm\,s}}{1-c_0\,e^{-2i\nu_\pm\,s}}}\quad\text{or}\quad \boxed{A_\pm(t)=\nu_\pm\,\frac{m\,\omega_{s}}{\hbar}\,\frac{1+c_0\,e^{-2i\nu_\pm\,\omega_{s}t}}{1-c_0\,e^{-2i\nu_\pm\,\omega_{s}t}}} \label{eqapm}
\end{equation}
with
\begin{equation}
\nu_\pm=\left\{ \begin{array}{rcl}
\small &1&\quad\text{if}\quad \vert \langle  z \rangle_+ -\langle  z \rangle_-\vert \leq 2R \\
&\vert \beta_\pm\vert&\quad\text{otherwise.}
\end{array}\right.\label{otherwise.}
\end{equation}
\\
Therefore, it is not necessary to solve the equations \eqref{gausg} for knowing $B_\pm(t)$, of which the values can be derived directly from \eqref{ehrenf} and using the analytical expressions of $A_\pm(t)$, $\langle  z \rangle_\pm$ and $\langle  p \rangle_\pm$, the latter being obtained making use of Ehrenfest's theorem.
\\ \\
The real part of equation \eqref{eqapm} can be rewritten in terms of the initial spread $\mathcal{Q}_0=\langle {\bf z}(t_0)^2\rangle -\langle {\bf z}(t_0)\rangle ^2$ as
\begin{equation}\label{reat}
\frac{1}{\mathcal{R}e\left(A_\pm(t)\right)}=2 \,\mathcal{Q}_0
   \cos ^2\left(\nu_\pm\,  \omega _s\,t\right)+\frac{\hbar ^2 \sin ^2\left(\nu_\pm  \, \omega _s\,t\right)}{2 m^2 \omega _s^2\,\nu_\pm ^2 \,\mathcal{Q}_0 }
\end{equation}
Hence since $\mathcal{Q}_\pm(t)=\frac{1}{2\,\mathcal{R}e\left(A_\pm(t)\right)}$ we have :
\begin{equation}\label{qpmt}
\mathcal{Q}_\pm(t)= \,\mathcal{Q}_0
   \cos ^2\left(\nu_\pm\,  \omega _s\,t\right)+\frac{\hbar ^2 \sin ^2\left(\nu_\pm  \, \omega _s\,t\right)}{4 m^2 \omega _s^2\,\nu_\pm ^2 \,\mathcal{Q}_0 }
\end{equation}
Note that the case of the free particle is recovered in the limit $\omega _s\,t\ll 1$ in which the equation above is expanded as
\begin{equation}
\mathcal{Q}_\pm(t)=\mathcal{Q}_0+\mathcal{Q}_0 t^2 \left(\frac{\hbar ^2}{4 m^2 \mathcal{Q}_0^2}-\nu_\pm ^2 \omega _s^2\right)+\mathcal{O}\left((\omega _s\,t)^3\right)
\end{equation}
which can be put into the form 
\begin{equation}\label{freeq}
\mathcal{Q}_\pm(t)=\underbrace{\mathcal{Q}_0\left[1+ \frac{\hbar ^2\,t^2}{4 m^2 \mathcal{Q}_0^2}\right]}_\textrm{Quantum spread of the free particle\small} -\underbrace{(\omega _s\,t)^2\,\mathcal{Q}_0\,\nu_\pm ^2}_\textrm{Contribution of self-gravity}+\mathcal{O}\left((\omega _s\,t)^3\right)
\end{equation}

\subsection{\bf Discussion:  contributions of self-gravity to the spread and nuclear corrections.}\label{nuclear}
In order to take account of nuclear corrections, we have to impose that $\omega_{s}$ is of the order of  $\sqrt{G\,({10^{-12} \over 10^{-10}})^{3}\rho_{sphere}}\approx 1$ Hz when the width of the wave function is smaller than the size of a nucleon (of the order of $10^{-12}$ m \cite{chen}), and equal to $\sqrt{\frac{G\,m}{R^3}}=\sqrt{G\,\rho_{sphere}}$ otherwise \cite{CDW}.

Even if we take these corrections into account, the global contribution of self-gravity in (\ref{freeq}) is actually negligible because either the pulsation of the (comoving) harmonic potential $\omega_{s}$ is very small or, when it is large due to nuclear contributions, it is large only during very short times.

Indeed it is easy to check that whenever the width of the wave function is larger than or equal to $10^{-10}$ m \cite{CDW}, the contribution of the nucleons vanishes and  $\omega _s\approx\sqrt{\frac{G\,m}{R^3}}=\sqrt{G\,\rho_{sphere}}\approx  10^{-3}$ Hz then $(\omega _s\,t)^2 \approx  10^{-6}$ for falls of a duration of the order of 1 s.

Otherwise, when for instance $\sqrt \mathcal{Q}_0\leq 10^{-12}$ m, the expansion of the free packet occurs so fast that the pulsation $\omega_{s}$ will get boosted by nucleic contributions during a negligibly short time. For instance, for a mass $m$ of $10^{-14}$ kg, if we impose that $\sqrt \mathcal{Q}_0$ belongs to the interval $(10^{-13},10^{-15})$  m, $\sqrt \mathcal{Q}_0$ will reach a width $\sqrt \mathcal{Q}\approx 10^{-10}$ m after a time of the order of $(m\sqrt \mathcal{Q}_0/\hbar)\cdot 10^{-10}$ m that belongs to the interval $(10^{-3},10^{-5})$ s. During this very short time, $\omega _s\,t$ will remain very small compared to unity. This explains why  the nucleic corrections are negligible being given the typical values of times, masses and so on considered by us in our proposal.

\begin{figure}\begin{center}\includegraphics[scale=0.67]{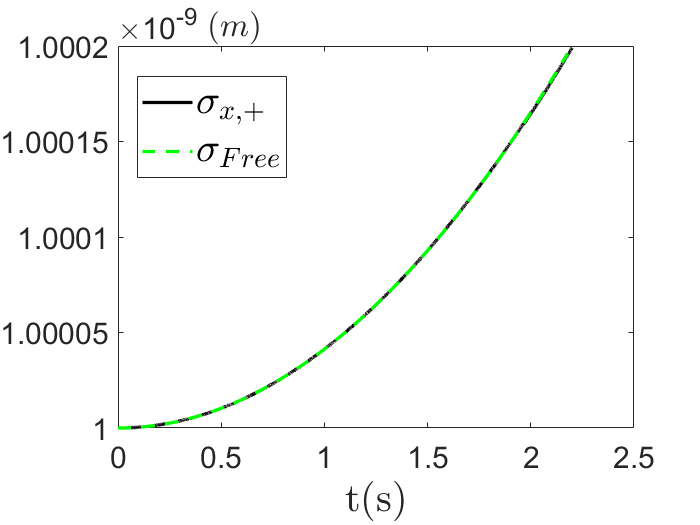}\caption{ Here we plot the spread in position of the wave packets $\ket{+}$, $\sqrt{\mathcal{Q}_{+}}=\frac{1}{\sqrt{2\,\mathcal{R}\text{e}A_+}}$ with and without self-gravity.  We considered $m=5,5.10^{-15}$ kg, $\omega_s=\sqrt{\frac{G\,m}{R^3}}\sim 6.10^{-4}$ Hz, $\vert \beta_+\vert=\frac{1}{\sqrt{3}}$ and $\vert \beta_-\vert=\sqrt{\frac{2}{3}}$ and $\sqrt{\mathcal{Q}_0}=10^{-9}$ m.}\label{aaaaa}\end{center}\end{figure}

As has been confirmed by accurate numerical computations (see e.g. figure \ref{aaaaa}), the difference between $\mathcal{Q}_\pm(t)$ and their free counterpart $\mathcal{Q}_0\left[1+ \frac{\hbar ^2\,t^2}{4 m^2 \mathcal{Q}_0^2}\right]$ can thus consistently be neglected.

 Indeed, in the limit $\omega _s\,t\ll 1$ considered above and using equation \eqref{freeq} we get that the difference in the width is given by \\
\begin{equation}\label{difreeq}
\mathcal{D}(t)=\sqrt{\mathcal{Q}_-(t)}-\sqrt{\mathcal{Q}_+(t)}=\frac{\sqrt{\mathcal{Q}_0}}{2}\, \left(1-2 \,\vert \beta_-\vert^2\right)\, (\omega _s\,t)^2+\mathcal{O}\left((\omega _s\,t)^3\right)
\end{equation}
\\
For example, in our simulations, when we considered $m=5,5.10^{-15}$ kg, $\omega_s=\sqrt{\frac{G\,m}{R^3}}\sim 6.10^{-4}$ Hz, $\vert \beta_+\vert=\frac{1}{\sqrt{3}}$ and $\vert \beta_-\vert=\sqrt{\frac{2}{3}}$, we found for a time $T_5$ of the order of $3$ s, $\mathcal{D}(T_5)\sim 5,4.10^{-17}$ m. The difference between the width $\mathcal{D}(T_5)$ is thus smaller than all the characteristic lengths considered in this study. Actually, the difference between $A_\pm(t)$ and $A(t)$, where $A(t)$ is the free counterpart of $A_\pm(t)$ (see also figure \ref{aaaaa}) can also be consistently neglected for similar reasons.
All this explains why 

\begin{align}\nonumber \frac{1}{\hbar}\int\,dt~\mathcal{F}_{Q,\pm}&=&\frac{1}{\hbar}\int\,dt(\frac{\hbar^2}{4m\mathcal{Q}_\pm}+\frac{m\omega_{s}^2}{2}\,\mathcal{Q}_\pm(t)\,\nu_\pm^2-\frac{6}{5}\,\frac{G\,m^2}{R}\,\nu_\pm^2-(1-\nu_\pm^2)\,\frac{Gm^2}{d})\\ &\approx& \frac{1}{\hbar}\int\,dt(\frac{\hbar^2}{4m\mathcal{Q}}+\frac{m\omega_{s}^2}{2}\,\mathcal{Q}(t)\,\nu_\pm^2-\frac{6}{5}\,\frac{G\,m^2}{R}\,\nu_\pm^2-(1-\nu_\pm^2)\,\frac{Gm^2}{d})\end{align}

\subsection{Formal expressions of the integrals contributing to the phase shift \label{diantrebis}}

We shall now consider the following integrals that appear in the total phase shift (\ref{diantre}):
\begin{align}
I_{1,\pm}(t)=\int~dt'\,\frac{\hbar}{4m\mathcal{Q}_\pm(t')} \quad\text{and}\quad I_{2,\pm}(t)=\int~dt'\,\frac{m\omega_{s}^2}{2\,\hbar}\,\mathcal{Q}_\pm(t')\,\nu_\pm^2.
\end{align}
$\bullet\quad$ It can be shown that 
\begin{equation}\label{i1dex}
I_{1,\pm}(t)=\frac{1}{2} \tan ^{-1}\left(\frac{\hbar \,\tan \left(\nu_\pm  \omega _s\,t\right)}{2 m \nu_\pm  \mathcal{Q}_0 \omega
   _s}\right)
\end{equation}
In the limit $\omega _s\,t\ll 1$ (this is the limit we considered in the main paper):
\begin{equation}
I_{1,\pm}(t)=\frac{\hbar t  }{4 m \mathcal{Q}_0}+\mathcal{O}\left((\omega _s\,t)^3\right).
\end{equation}
In this limit, self-gravity does not contribute to the phase shift since $I_{1,+}(t)-I_{1,-}(t)=0+\mathcal{O}\left((\omega _s\,t)^3\right)$.
\\

\begin{figure}
\begin{center}
	\includegraphics[scale=0.57]{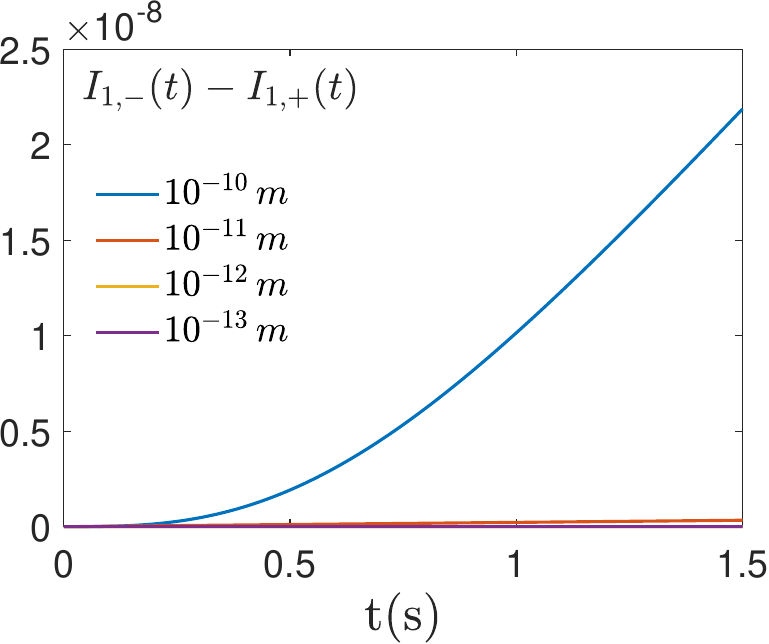}
	\includegraphics[scale=0.57]{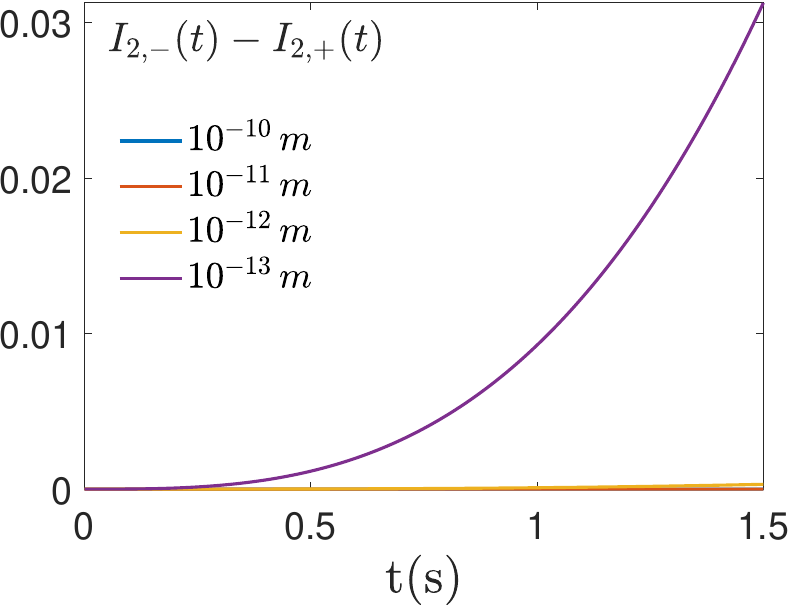}
	\caption{We illustrate here the functions \eqref{i1dex} and \eqref{i2dex} in the limit $\omega _s\,t << 1$ for different values of the initial spread $\sqrt{\mathcal{Q}_0}$. We chose $m=5,5.10^{-15}$ kg, $\omega_s=\sqrt{\frac{G\,m}{R^3}}\sim 6.10^{-4}$ Hz, $\vert \beta_+\vert=\frac{1}{\sqrt{3}}$ and $\vert \beta_-\vert=\sqrt{\frac{2}{3}}$}\label{i12d}
\end{center}
\end{figure}
$\bullet\quad$ Now let us consider the second contribution made by the function $I_{2,\pm}(t)$. It can be shown that after integration we get:
\\
\begin{equation}\label{i2dex}
I_{2,\pm}(t)=\frac{1}{2} \,\frac{m \omega _s^2}{\hbar}\,\nu_\pm ^2\, \left[\frac{\mathcal{Q}_0 t}{2}+\frac{\hbar^2\,t  }{8 m^2\omega _s^2 \nu_\pm ^2 \,\mathcal{Q}_0 }+\left(\frac{\mathcal{Q}_0}{4 \nu_\pm  \omega _s}-\frac{\hbar ^2 }{16 m^2 \nu_\pm ^3 \omega _s^3\,\mathcal{Q}_0
   }\right)\,\sin \left(2 \nu_\pm  \omega _s\,t\right)\right]
\end{equation}
\\
 Here again, in the range of parameters (initial spreads, mass, typical times) considered by us, $\omega _s\,t\ll 1$ and $1\,\ll  \frac{\hbar ^2\,t^2}{4 m^2 \mathcal{Q}_0^2}$ excepted for very short times so that 
 \begin{equation}
I_{2,\pm}(t)\approx \frac{1}{2} \,\frac{m \omega _s^2}{\hbar}\,\nu_\pm ^2\,\mathcal{Q}_0\left[ \frac{\hbar ^2\,t^3}{12 m^2 \mathcal{Q}_0^2}\right]
\end{equation}
Hence if $\vert \beta_+\vert=\frac{1}{\sqrt{3}}$ and $\vert \beta_-\vert=\sqrt{\frac{2}{3}}$:
 \begin{equation}\label{q0int}
I_{2,+}(t)-I_{2,-}(t)\approx \frac{m \omega _s^2}{\hbar}\mathcal{Q}_0\left[ \frac{\hbar ^2\,t^3}{72 m^2 \mathcal{Q}_0^2}\right]
\end{equation}
which fits well with the analytical function plotted in figure \ref{contf}.

In particular, if one can initially prepare the degrees of freedom associated with the center of mass (or at least their $z$ component) in the ground state of the trap from which it is afterward released, then $\mathcal{Q}_0=\hbar/(m\,\omega_{Trap})$,  and the contribution to the phase-shift reads
 \begin{equation}\label{q0int2}
I_{2,+}(T_5)-I_{2,-}(T_5)\approx\frac{1}{72}\,\omega_{Trap}\,\omega_{s}^2\,\left(T_5-T_s\right)^3
\end{equation}
for $T_5$ sufficiently large. 
For instance in equation \eqref{q0int}, if we take $m=5,5.10^{-15}$ kg, $\sqrt{\mathcal{Q}_0} = 10^{-13}$ m, we predict $\omega_{Trap}=1.82$ Mhz and $\omega_{s}=6,4.10^{-4}$ Hz. Imposing $T_5=2$ s and $T_s=0.034$ s, the contribution in the phase shift made by $I_{2,+}(T_5-T_s)-I_{2,-}(T_5-T_s)$ is predicted in this way to be of the order of $0.07039$ which fits very well the exact final phase shift $\approx 0.07035$ corresponding to (the lower left plot in) figure \ref{contf}. In figure \ref{i12d} we plot the functions  $I_{1,+}(t)-I_{1,-}(t)$ and $I_{2,+}(t)-I_{2,-}(t)$ for different values of the initial spread. They correspond to the upper and lower left plots of figures \ref{conte} and \ref{contf} respectively.

\tom{\subsection{Entangling power and structure of the interaction.\label{appendpower}} It has been shown elsewhere by one of us \cite{zeit} that the entangling power of an interaction between two systems $A$ and $B$ initially prepared in a factorizable state is proportional to the coupling rate of the interaction Hamiltonian to states biorthogonal to this factorizable state. In the same paper, the following theorem has been shown: if a specific Hamiltonian $H_{AB}$ never entangles the systems $A$ and $B$ initially prepared in an arbitrary factorizable state, then $H_{AB}=H_A1\!\!1_B+1\!\!1_AH_B$, and there is no effective interaction between $A$ and $B$ (only effective self-interaction). Now, the interaction Hamiltonian between the systems $A$ and $B$ in a double Stern-Gerlach device is obviously the sum of the self-interactions between each system plus the mean field interaction generated by each system on its partner. It is easy to show that it reads $H_{AB}=H^{self}_A1\!\!1_B+1\!\!1_AH^{self}_B+H_A^{<B>}1\!\!1_B+1\!\!1_AH^{<A>}_B$, where $H^{self}_{A(B)}$ as well as $H_A^{<B>}$ ($H_B^{<A>}$) are a linear combination (with real amplitudes) of $1\!\!1_{A(B)}$ and $\sigma^Z_{A(B)}$=$\ket{+^{A(B)}}\bra{+^{A(B)}}-\ket{-^{A(B)}}\bra{-^{A(B)}}$. This means that neither the self-interaction nor the effective interaction resulting from the mean influence of the other system will make it possible to entangle these systems. In a sense, this is a contraposition of another theorem, valid in the linear regime, shown in reference \cite{zeit}: if the total system factorizes, each system feels (up to a global phase) the mean influence of the other one \cite{zeit,gemmer}. However, here, each system ``feels'' the mean influence of the other one, even if the full state is entangled, which is a signature of the S-N interaction, in full conformity with the mean-field approach initiated by M{\o}ller \cite{Moeller} and Rosenfeld \cite{Rosenfeld}.}

\end{document}

%% file: main.bbl
%